\documentclass[conference,compsoc]{IEEEtran}
\IEEEoverridecommandlockouts
\makeatletter
\renewcommand{\@IEEEsectpunct}{~}
\makeatother
\usepackage{cite}
\usepackage{amsmath,amssymb,amsfonts}
\usepackage{algorithmic}
\usepackage{graphicx}
\usepackage{textcomp}
\usepackage{xcolor}
\usepackage{makecell}
\usepackage{lipsum}
\usepackage{xspace}
\usepackage{tikz}
\usepackage{comment}
\usepackage{enumitem}
\usepackage{array}
\usepackage{chemformula}
\usepackage{nicematrix}
\usepackage{pgfgantt}
\usepackage{rotating}
\usepackage{array,multirow,graphicx}
\usepackage{float}
\usepackage{cleveref}
\usepackage{subcaption}
\usepackage{url}
\usepackage{tabularx,booktabs}

\makeatletter 
\newcommand{\linebreakand}{%
  \end{@IEEEauthorhalign}
  \hfill\mbox{}\par
  \mbox{}\hfill\begin{@IEEEauthorhalign}
}
\makeatother 

\newcolumntype{P}[1]{>{\raggedright\vrule height4ex width 0pt}p{#1}<{\vrule depth 2.5ex width 0pt}}

\definecolor{notecolor}{rgb}{0.75,0,0}

\newcommand{\burnindesign}{\textit{Target}\xspace}

\begin{document}

\title{Pentimento: Data Remanence in Cloud FPGAs}

\author{

\IEEEauthorblockN{Colin Drewes}
\IEEEauthorblockA{
\textit{Stanford University} \\
\textit{University of California San Diego} \\
drewes@cs.stanford.edu}\\

\IEEEauthorblockN{Alric Althoff}
\IEEEauthorblockA{
\textit{ARM} \\
alric.althoff@arm.com
}

\and

\IEEEauthorblockN{Olivia Weng}
\IEEEauthorblockA{
\textit{University of California San Diego} \\
oweng@ucsd.edu
}\\ \\

\IEEEauthorblockN{David Kohlbrenner}
\IEEEauthorblockA{
\textit{University of Washington} \\
dkohlbre@cs.washington.edu
}

\and

\IEEEauthorblockN{Andres Meza}
\IEEEauthorblockA{
\textit{University of California San Diego} \\
anmeza@ucsd.edu
}\\ \\ 

\IEEEauthorblockN{Ryan Kastner}
\IEEEauthorblockA{
\textit{University of California San Diego} \\
kastner@ucsd.edu
}

\linebreakand

\IEEEauthorblockN{Dustin Richmond}
\IEEEauthorblockA{
\textit{University of California, Santa Cruz} \\
drichmond@ucsc.edu
}
}

\maketitle

\begin{abstract}
Cloud FPGAs strike an alluring balance between computational efficiency, energy efficiency, and cost.  
It is the flexibility of the FPGA architecture that enables these benefits, but that very same flexibility that exposes new security vulnerabilities. We show that a remote attacker can recover ``FPGA pentimenti'' -- long-removed secret data belonging to a prior user of a cloud FPGA.
The sensitive data constituting an FPGA pentimento is an analog imprint from bias temperature instability (BTI) effects on the underlying transistors. We demonstrate how this slight degradation can be measured using a time-to-digital (TDC) converter when an adversary programs one into the target cloud FPGA.

This technique allows an attacker to ascertain previously safe information on cloud FPGAs, even after it is no longer explicitly present. Notably, it can allow an attacker who knows a non-secret ``skeleton'' (the physical structure, but not the contents) of the victim's design to (1) extract proprietary details from an encrypted FPGA design image available on the AWS marketplace and (2) recover data loaded at runtime by a previous user of a cloud FPGA using a known design. Our experiments show that BTI degradation (burn-in) and recovery are measurable and constitute a security threat to commercial cloud FPGAs.

\end{abstract}
\begin{figure*}[!htb]
    \begin{subfigure}[t]{1\linewidth}
        \centering
        \includegraphics[width=\linewidth]{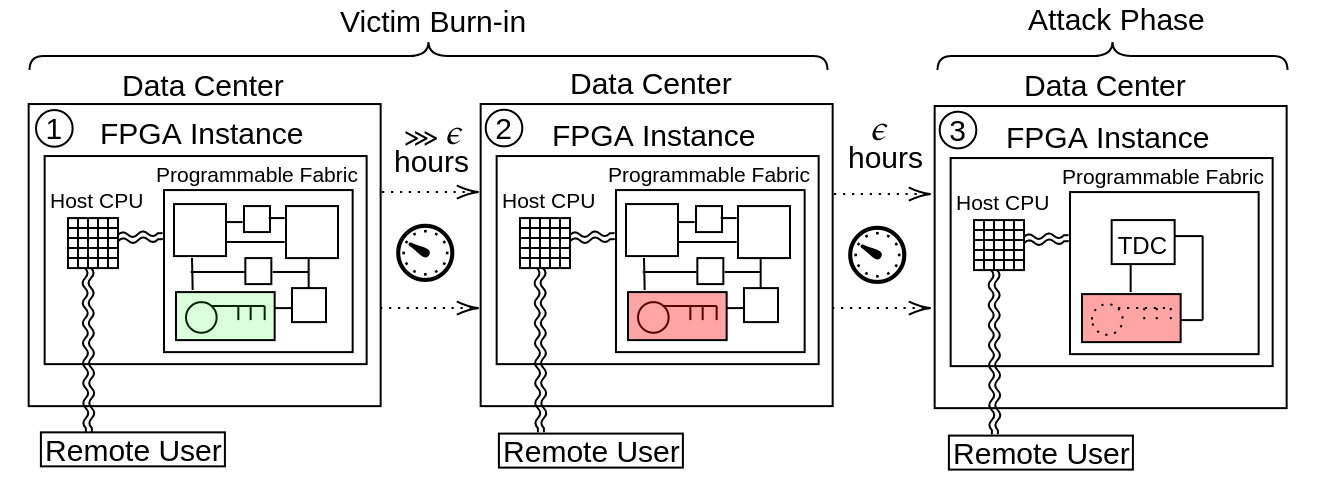}
    \end{subfigure}
    \vspace{-15px}
    \caption{\textbf{Pentimenti FPGA Threat Models:} \protect\textcircled{\raisebox{-.8pt} {1}} A design containing confidential information (green key) is loaded onto the FPGA. \protect\textcircled{\raisebox{-.8pt} {2}} After this design runs for some number of hours, parts of the design are imprinted -- a pentimento (red key) is left on the FPGA due to analog remanence from BTI effects. \textcircled{\raisebox{-.8pt} {3}} The attacker loads their design with a BTI sensor to extract the pentimento based on the recovery timing effects. 
    }
    \label{fig/threat_model}
    \vspace{-15px}
\end{figure*}
\section{Introduction}
Amazon,  Microsoft, Alibaba, Baidu, Huawei, TenCent, and Nimbix offer FPGAs as an on-demand cloud service. FPGAs efficiently accelerate common cloud applications including neural networks~\cite{fowers2018configurable}, video transcoding~\cite{fpga-transcoding}, genome sequencing~\cite{chen2016spark}, secure database transactions~\cite{arasu2015transaction}, networking~\cite{putnam2014reconfigurable}, and homomorphic encryption~\cite{poppelmann2015accelerating}.

Unfortunately, cloud FPGAs open the door to new 
security vulnerabilities related to confidentiality~\cite{schellenberg2018inside, zhao2018fpga, schellenberg2018remote, glamovcanin2020cloud, giechaskiel2019reading}, integrity~\cite{mahmoud2019timing, krautter2018fpgahammer, boutros2020neighbors, rakin2021deep, luo2021deepstrike}, and availability~~\cite{gnad2017voltage, matas2020power}. Signal timing sensors~\cite{zick2012low, gnad2016analysis} have been used to extract cryptographic keys of active computation within the FPGA~\cite{schellenberg2018inside}, identify the active computation running within the FPGA~\cite{gobulukoglu2021dac}, implement covert channels across dies on a 2.5D integrated package~\cite{giechaskiel2019reading}, and perform attacks across chips on the same board~\cite{schellenberg2018remote, giechaskiel2020c3apsule}. 

These attacks require that the attacker and victim are spatiotemporally co-located on the same system. 
For this reason, cloud FPGAs are often only temporally shared; they do not allow multiple users to co-exist spatially on the same FPGA. Conventional wisdom says information leakage will not occur if the FPGA is correctly erased after use.
Thus, after a user relinquishes the cloud FPGA, the FPGA is wiped~\cite{leeser2021fpgas, aws-fpga-faq}, and at some point, rented to another user.

The attacks presented in this work exploit a side channel that allows an attacker to target previous users of the FPGA even after a wiping procedure is performed. The victim is no longer performing computation or renting the FPGA; the victim no longer resides on the device, \textit{and the victim has left no logical information on the device.}  

We show that data from previous users can be extracted via an analog side channel due to bias temperature instability (BTI) aka \emph{burn-in}.
We call these \emph{``FPGA pentimenti''} -- the analog residue of the previous digital data that remains on the FPGA due to BTI effects. FPGA pentimenti are recoverable by sensing BTI recovery using time-to-digital (TDC) sensors. Our experiments show that FPGA pentimenti are a real and extant threat to the security of cloud FPGAs. Much like infrared imaging can expose art work pentimenti (previous paint strokes since painted over by an artist) not visible to the naked eye, we demonstrate that attackers can exploit \textit{FPGA pentimenti} (previous design and user data digitally wiped by the cloud provider).

BTI physically deteriorates transistors, thus negatively affecting their propagation delay.
The BTI effect is caused by applying positive/negative (1/0) voltages to CMOS transistors. 
\emph{BTI recovery} occurs when the transistors are no longer stressed; the transistors partially revert to their previously faster state. Transistors undergo negative and positive BTI caused by applying logical $0$ and $1$ values, respectively. NBTI and PBTI degradation is not symmetric. NBTI effects are typically larger than PBTI. NBTI recovery is also faster. By measuring the speed and size of the recovery, an attacker can deduce if a previous value as a $1$ or $0$.

We show how to measure BTI effects on cloud FPGAs to exploit an  analog temporal side channel that leaks data between successive users of a cloud FPGA.
BTI is measured using a TDC sensor that records the time to propagate a pulse through FPGA resources.
The change in propagation delay reflects the previous values held by those resources due to BTI effects. We show how this opens a side channel that can be exploited by an attacker to ascertain previous design and user data.

Our work describes, for the first time, how BTI effects are a security concern for commercial cloud FPGAs. We demonstrate the ability to recover pentimenti in remote FPGAs to expose two violations of the cloud FPGA security model when the ``skeleton'' (the physical structure, but not the contents) of the design is known to the attacker: An attacker can (1) extract proprietary details or keys from an encrypted bitstream accessible via the cloud platform (i.e., the AWS marketplace) and (2) recover non-transient runtime data from a previous user of a cloud FPGA device by observing the BTI recovery via circuit timing changes.

We experimentally validate the burn-in threat on a AWS F1 (Virtex UltraScale+) cloud FPGA and on a local ZCU102 (Zynq UltraScale+) FPGA.
In both cases, we demonstrate a discernible difference of the burn-in behavior on an FPGA route before and after BTI degradation.
The timing behavior of that route is dependent on what the previous value was, and thus an attacker is able to carry out attack (1) and (2) on the AWS F1 platform as detailed in Section~\ref{section/threat_model}. 

We contextualize these findings on the OpenTitan hardware root of trust -- an open-source hardware design with strict data security requirements~\cite{OpenTitan}.
Roots-of-trust carry out core, security-critical functionalities such as secure boot, configuration of operative modes (e.g., debug vs normal), and management of sensitive of data (e.g., cryptographic keys).
OpenTitan security assets are vulnerable to the burn-in threat.

Section~\ref{section/threat_model} presents the cloud FPGA threat model.
Section~\ref{section/transistor_degradation} provides background on the effects of BTI transistor degradation, which are measured by the sensor presented in Section~\ref{section/sensor_structure}.
We construct an experiment to verify the burn-in threat model generally on an Ultrascale+ FPGA and specifically on the AWS F1 platform in Section~\ref{section/experimental_design}, which is carried out in Section~\ref{section/experimental_results}.
We relate this paper to prior efforts in Section~\ref{section/related_work}, discuss mitigations in Section~\ref{section/mitigations}, and conclude in Section~\ref{section/conclusion}. 

\textbf{Disclosure:} 
The necessary steps have been taken to alert affected vendors.
Amazon Web Services was originally notified July 2022. 
Xilinx was originally notified in August 2022.
\section{Threat Models}
\label{section/threat_model}

The threat models extract side channel information about previous cloud FPGA user data   via temporal analog residue, aka ``pentimenti'', that arise from BTI recovery effects. Our discussion is framed in the context of the AWS F1 platform though it applies to other cloud FPGA platforms.
AWS enables customers to share/sell preexisting designs to other AWS users through the AWS marketplace.
AWS provides these designs as an Amazon Machine Image (AMI) and Amazon FPGA Image (AFI).
The AFI provides the FPGA bitstream.
The AMI is a Linux machine that interfaces with the AFI. 

Figure~\ref{fig/threat_model} shows the general approach of our threat models. \textcircled{\raisebox{-.8pt} {1}} A user rents and loads a design containing confidential information (denoted by the green key). \textcircled{\raisebox{-.8pt} {2}} The design remains programmed on the FPGA and computes for some number of hours, allowing the user data to experience BTI effects and burn-in (red key).  
The victim FPGA is released back into the rental pool. It undergoes a design wipe performed by AWS to reset the system and clear out any data remanence~\cite{leeser2021fpgas, aws-fpga-faq}. \textcircled{\raisebox{-.8pt} {3}} The attacker gains access to the FPGA and loads the TDC sensor to extract the pentimenti -- the analog residues of the previous digital data that remains on the FPGA due to BTI effects.

With this setup, we can extract two types of previously safe data using the techniques presented in this paper: \textbf{Type A} design data and \textbf{Type B} user data.

\textbf{Type A (Design Data):} FPGA designs often contain confidential information as netlist constants, e.g., cryptographic keys or machine learning weights.
The AFI promises to keep such proprietary design information secret. 
A purchased AFI does not permit the user access to the FPGA source code or bitstream\footnote{The bitstream is the binary file used to program the FPGA.} to preserve intellectual property rights. But this sensitive information can be extracted via their pentimenti as we show in this paper.
We call this secret information baked into the design \textbf{Type A} data, and the victim is the AFI publisher.
 
\textbf{Type B (User Data):}
Type B data is from a previous user of the FPGA. The previous user loads confidential information onto an AFI at runtime. Since the attacker does not control the loading and unloading of the design, an attack cannot rely on gathering initial delay estimates (as can be done for Type A data). Thus, extracting Type B user data is more challenging but powerful attack that requires measuring BTI recovery. 

\textit{The difference between \textbf{Type A} and \textbf{Type B} is subtle but shifts the target of an attack from being the publisher of a design/AFI (\textbf{Type A}) to the user of design/AFI (\textbf{Type B}).} The threat models differ, but both follow the steps depicted in Figure~\ref{fig/threat_model}.

\textbf{Threat Model 1 - Proprietary Design Data Extraction:}  Threat Model 1 targets Type A Design Data encoded into the design itself, e.g., a netlist constant holding a cryptographic key or machine learning weight. The attacker is renting the design, satisfies Assumptions 1 and 2 (discussed later), and can control the loading and unloading of the design. AWS guarantees to keep design intellectual property secret~\cite{aws-fpga-faq}, thus Threat Model 1 violates AWS F1 security guarantees.

The attacker extracts proprietary design information via the following: 
\begin{enumerate}
    \item A malicious AWS F1 user rents an FPGA instance with the intention extracting sensitive information from a third party design. 
    \item The attacker measures the routes that will hold the sensitive data and gather pre-burn-in route delay characteristics. 
    \item The attacker loads a target design in Stage \textcircled{\raisebox{-.8pt} {1}} of Figure~\ref{fig/threat_model} that contains sensitive information stored in the FPGA routes.
    \item The attacker executes the target design until Stage \textcircled{\raisebox{-.8pt} {2}} of Figure~\ref{fig/threat_model} when the BTI effects burn-in the FPGA routes holding sensitive information. 
    
    \item The attacker initiates the attack phase (Stage \textcircled{\raisebox{-.8pt} {3}} of Figure~\ref{fig/threat_model}). They unload the victim design and load a measure design that contains the TDC sensor from Section~\ref{section/sensor_structure} to measure the BTI degradation of the victim routes via their timing behavior. 
    
    \item The attacker analyzes sensor data to determine the sensitive information from the victim design with high probability. 
\end{enumerate}

\textbf{Threat Model 2 - Confidential User Data Extraction:}
The attacker recovers confidential data from a previous victim tenant of the cloud FPGA (Type B).
This model assumes that the attacker can requisition an FPGA after the victim has finished computing. 
The attacker extracts confidential user data via the following: 
\begin{enumerate}
    \item A non-malicious AWS F1 victim user loads a design in Stage \textcircled{\raisebox{-.8pt} {1}} of Figure~\ref{fig/threat_model}. This design contains sensitive information either stored statically in FPGA bitstream (e.g., a netlist constant) or in data loaded at runtime.
    
    \item  The victim design executes, during which the sensitive data is statically held in the FPGA resources. After some time, the victim design has induced the burn-in effect (Stage \textcircled{\raisebox{-.8pt} {2}} of Figure~\ref{fig/threat_model}). 
    
    \item The victim completes their computation and relinquishes the FPGA back into AWS’s pool of available devices.

    \item The attacker instantiates an AWS instance and is assigned the relinquished victim device.

    \item The attacker loads in Stage \textcircled{\raisebox{-.8pt} {3}} of Figure~\ref{fig/threat_model} an FPGA design that contains TDC sensors connected to victim resources that previously held sensitive information. 

    \item The attacker analyzes TDC sensor results to determine the sensitive victim data  with high probability.
\end{enumerate}

The difference between these two threat models shifts the attack target from the producers of the design IP  (\textbf{Threat Model 1}) to a previous user the cloud FPGA (\textbf{Threat Model 2}).
Both threat models are a fundamental violation of the AWS FPGA F1 security guarantees. AWS guarantees that ``no FPGA internal design code is exposed''~\cite{aws-fpga-faq} through an AFI leased from the marketplace, meaning \textbf{Threat Model 1} should not occur. Furthermore, AWS states that they scrub ``FPGA state on termination of an F1 instance,''~\cite{aws-fpga-faq} meaning \textbf{Threat Model 2} leakage should not occur. Our results demonstrate the feasibility of these threat models, which show that burn-in is recoverable using a TDC sensor. 

Our threat models rely on two assumptions. Assumption 1 covers Type A data and Threat Model 1. Type B data and Threat Model 2 make an additional assumption.

\textbf{Assumption 1:}
The attacker knows the placement, or ``skeleton'', of the targeted design routes\footnote{The wire segments inside the FPGA holding the targeted data.} that contains confidential design information (Type A) or  sensitive user data (Type B). 

The attacker's knowledge of the sensitive information's location could be derived from a publicly available design or bitstream. For example, the OpenTitan hardware root of trust distributes a prebuilt bitstream that a user loads with sensitive information like cryptographic keys~\cite{opentitan-fpga-setup}. Xilinx FINN provides prebuilt bitstreams for different neural network architectures~\cite{finn-examples}. In both cases, the complete source code and compilation scripts are available, which allows one to determine the locations of the sensitive data -- the keys for OpenTitan and the neural network weights for FINN. 

Other options to learn the target route places include: 1) the attacker is the original author of the AFI on the AWS marketplace and knows design route details. 2) Proprietary information about the design layout has been leaked to an attacker. Finally, when evaluating an implementation's security, it is common practice to assume the architecture is publicly visible~\cite{petitcolas2011kerckhoffs}. 

Thus, we believe it is reasonable to assume that the attacker knows the placement information (Assumption 1). Loosen or removing this assumption would strengthen the threat model, and we are considering ways to expand the threat model without Assumption 1 in future work. 

\textbf{Assumption 2: }The attacker has the ability to gain access to the same FPGA the victim relinquished. Gaining access to a relinquished cloud FPGA requires aspects of cloud cartography and co-location attacks~\cite{ristenpart2009hey, zhang2011homealone,aviram2010determinating, xu2015measurement, inci2016co} that check out devices en masse or leveraging cloud FPGA fingerprinting techniques~\cite{tian2019temporal, tian2020cloud, tian2020fingerprinting, tian2021cloud}. Another potential option is a flash attack where the attacker locks up the available stock right before the victim releases their instance. If the attacker procures all the available resources, they are guaranteed to obtain the relinquished victim board. In our AWS experimentation we commonly received errors implying that we have reached the limit of F1 devices in the region, suggesting that this flash attack could be accomplished through acquiring only a few devices.
\begin{figure}[!htb]
\centering
\includegraphics[width=\linewidth]{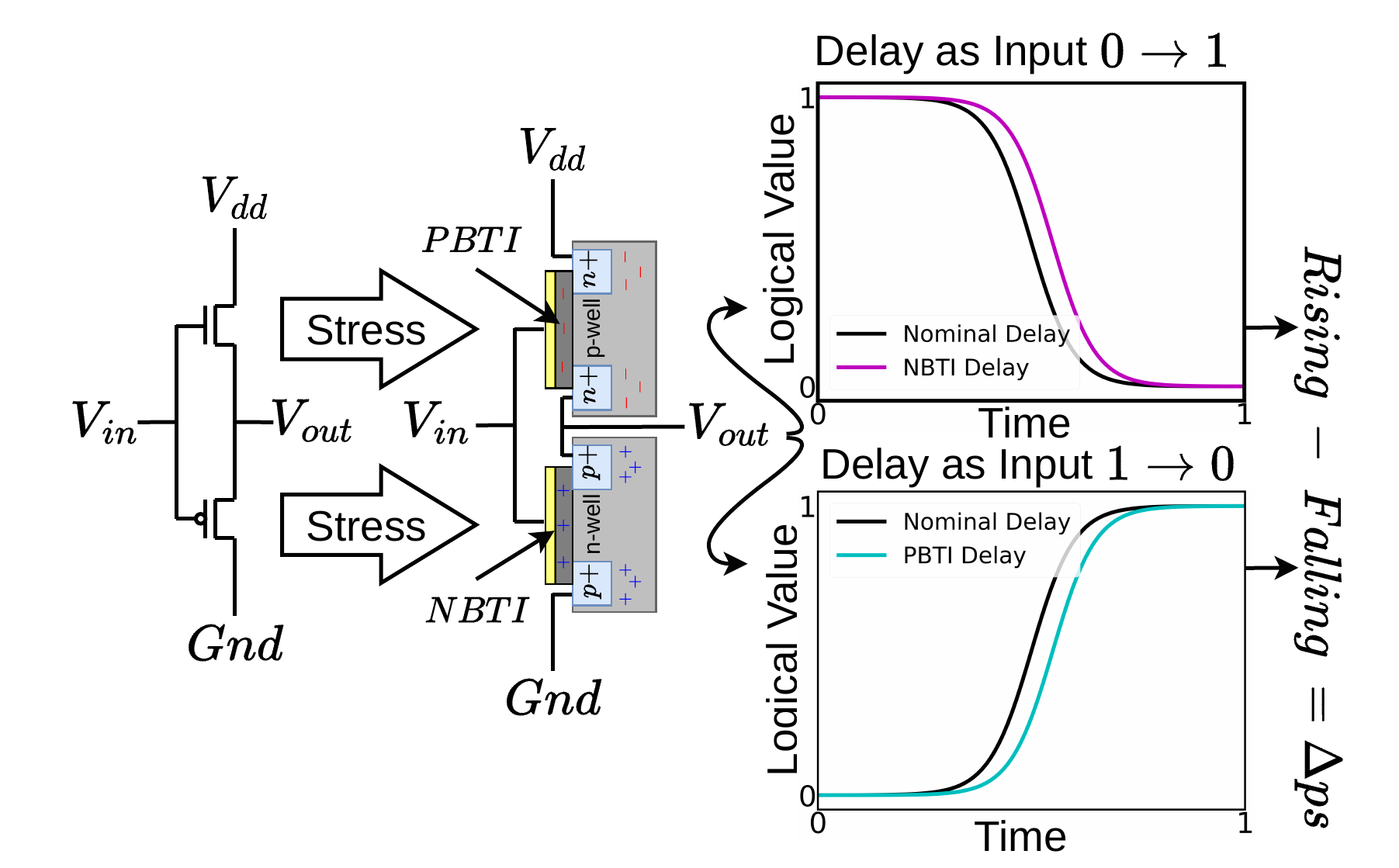}
\caption{Bias temperature instability (BTI) effects induce circuit delays that differ depending on the data computed. An inverter is pictured on the left, composed of a PMOS (bottom) and NMOS (top) transistor. A $V_{in}$ value of 0 (1) will allow current to flow through the PMOS (NMOS) transistor, causing it to degrade through NBTI (PBTI). As NBTI (PBTI) manifests, the speed in which the inverter stabilizes on an output, $V_{out}$, as $V_{in}$ equals 0 (1) will begin to slow. $\Delta$ps is the difference between the 0-input propagation speed (measuring PBTI) and the 1-input propagation speed (measuring NBTI). $\Delta$ps will vary depending on whether the inverter was previously computing a 0 or 1 value; thus, it can be used to infer the values of a previous computation.
}
\label{fig/bti_high_level}
\end{figure}

\section{Bias Temperature Instability (BTI)}
\label{section/transistor_degradation}

Bias temperature instability (BTI) is a degradation behavior of transistor transconductance, subthreshold slope, and linear and saturation drain current fundamental to modern field-effect transistors~\cite{mahapatra2013comparative, mahapatra2016fundamentals}.
\emph{Negative BTI (NBTI)} occurs when the PMOS transistor gate voltage is negative relative to its other terminals ($0$/\texttt{False} logical value), which results in positive charge migration into the silicon dioxide insulation.
\emph{Positive BTI (PBTI)} affects NMOS transistors when its gate voltage is positive relative to the other terminals ($1$/\texttt{True} logical value), resulting in negative charge migration into the insulating dielectric.
BTI effects accumulate under voltage stress, increasing the threshold voltage of their respective transistor types, and consequently increased transistor rise and fall transition delays~\cite{mangalagiri2008thermal}.

CMOS logic gates (e.g., NAND, NOR, and INV) are built from PMOS and NMOS transistors. Static 0/1 inputs on logic gates cause NBTI/PBTI degradation on PMOS/NMOS transistors. The degradation affects the rising (0 $\rightarrow$ 1) and falling (1 $\rightarrow$ 0) propagation delays through the logic gates. Figure~\ref{fig/bti_high_level} details how a CMOS inverter undergoes data-dependent BTI effects which cause timing deviations captured by the difference in rising and falling propagation delays through the inverter.

When BTI-causing values are removed, there is a partial threshold voltage recovery~\cite{chen2003dynamic, rangan2003universal, reisinger2006analysis, bhardwaj2006predictive, ramey2014bti} that increases the transistor switching speed.
NBTI and PBTI recovery differs in mechanism and recovery rate~\cite{hehenberger2010recovery, lee2013samsung16, grasser2007simultaneous}.
NBTI recovery is due to defect removal via the recuperation of broken bonds with positively-charged hydrogen atoms~\cite{krishnan2005material, ramey2014bti}; 
PBTI recovery is due to the removal of trapped negatively-charged electrons in the transistor dielectric~\cite{mahapatra2016fundamentals}. 
The defects from PBTI electron charge traps are energetically deeper than NBTI positive charge traps~\cite{zhang2022bias} and affect the recovery timescale of PBTI relative to NBTI.
BTI recovery effects are measurable, the differences between positive and negative recovery characteristics are apparent, and we can exploit them as a temporal side channel. 

BTI creates pentimenti -- analog remenance of previous design state and data. BTI effects are differentiable; they encode data about the prior state, e.g., if the route under test was previous a 0 or 1 value. The rate and degree of BTI effects are driven by constant voltages and dynamic switching~\cite{stott2010degradation}, and has a predictable effect. Remenance is an essential condition to recover any logical information  and necessary for Threat Models 1 and 2. BTI effects are not-permanent. 
BTI degradation is not a permanent artifact and undergoes a recovery period where the transistors become faster~\cite{chen2003dynamic, rangan2003universal, reisinger2006analysis, bhardwaj2006predictive, ramey2014bti}. 
This required for Threat Model 2 since it requires measuring remanance of previous user values that are no longer present.

FPGAs contain many resources that undergo BTI and can be targeted in pentimenti attacks: bitstream configuration bits, programmable routing, configurable logic blocks (CLBs), digital signal processors (DSPs), and block RAMs (BRAMs). 
In order to perform a successful attack, the victim resource should meet the following conditions:

\begin{itemize}
    \item \textbf{BTI effects must occur:} The target resource must be used in a manner that induces burn-in (BTI degradation) -- a necessary condition to recover Type A and B data. For example, a route is statically held at constant value. 
    \item \textbf{BTI effects must be differentiable:} The target resource should exhibit differences in circuit-level behavior due to BTI degradation and recovery. For example, the route delay profile differs based on whether it was previously held at logical 0 or 1.
    \item \textbf{BTI-effected resources must be observable:} Targeted resources must be in user-visible locations on AWS F1. Some cloud FPGA resources are inaccessible by the user, e.g., resources implementing the AWS shell. The attacker is limited by the interfaces exposed by the cloud provider. The attacker do not have physical access. They cannot use special sensing instrumentation. The BTI sensor must be implementable by any user, without elevated privilege, and pass design rule checks. 
\end{itemize}

\begin{figure*}[!htb]
\centering
\begin{subfigure}[t]{\linewidth}
  \centering
  \includegraphics[width=\linewidth]{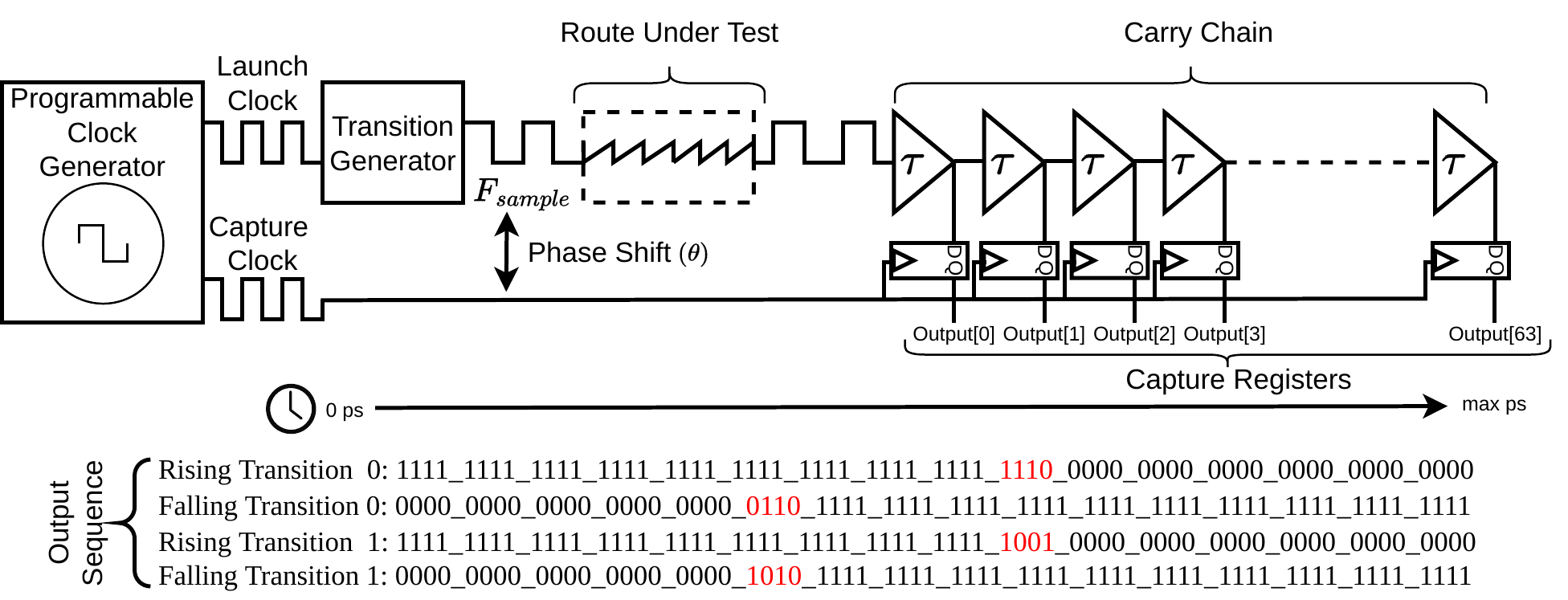}
\end{subfigure}
\caption{\textbf{Time-to-Digital Converter (TDC) sensor that measures BTI effects in FPGA routes.} A \texttt{Programmable Clock Generator} produces two clock domains: the \texttt{Launch Clock} and \texttt{Capture Clock}. The \texttt{Launch Clock} domain generates a signal that propagates through a test route and into a texttt{Carry Chain} -- an array of linear delay elements (64 in this example). The signal moves through the Route Under Test and into the TDC \texttt{Carry Chain}, when it is recorded as indicated by the \texttt{Capture Clock}. The TDC \texttt{Capture Registers} record the signal propagation distance. This provides is a measure of propagation delay of the \texttt{Route Under Test}. Taking measurements over time records changes of the propagation delay due BTI remnants stored on the transistors.}
\label{fig/sensor_architecture}
\vspace{-15px}
\end{figure*} 

FPGA programmable routing meets all three conditions.
Specifically, we target the route between an FPGA register and a CLB.
The programmable route can be composed to be arbitrarily long sequences of transistors to increase the observable BTI effects, and it is trivial to use as a route. 
Additionally, programmable routes often carry sensitive data (e.g., encryption keys and machine learning weights).
Thus, verifying that a route between a register and LUT is vulnerable to a pentimenti attack threatens the data integrity of most FPGA designs.

\section{BTI Sensor}
\label{section/sensor_structure}
Our threat models depend on the ability to detect BTI effects in cloud FPGA resources. BTI degradation and recovery manifests as changes in the timing delays of the victim resource.
BTI effects differ depending on the previous state that the victim held on that resource.

Time-to-digital converters (TDCs)  measure ns-scale timing changes by sensing the propagation delay through an FPGA-instantiable delay line.
There exists a large body of prior work on implementing TDC sensors within cloud FPGAs~\cite{schellenberg2018remote, schellenberg2018inside, gnad2021voltage, giechaskiel2020c3apsule, gobulukoglu2021dac, drewes2023turnon}. 
Our experiments use the open-source Tunable Dual-Polarity TDC~\cite{drewes2023turnon}.

Figure~\ref{fig/sensor_architecture} demonstrates how to use a Tunable Dual-Polarity TDC sensor~\cite{drewes2023turnon} to measure BTI effects.
The original sensor is designed for power measurement; we amend it to perform BTI timing measurement and exploit the threats presented in Section~\ref{section/threat_model}.  
The constituent structures of the TDC are presented below:

\subsubsection*{Programmable Clock Generator:}
This component generates the two clock domains: the \texttt{Launch Clock} and \texttt{Capture Clock}. These two clocks are identical in frequency with a runtime-programmable phase relationship defined by $\theta$. Two clocks are necessary as the TDC is comparing how long it takes for the signal from the \texttt{Launch Clock} to reach a destination within the \texttt{Carry Chain} compared to when the \texttt{Capture Clock} causes the \texttt{Capture Registers} to sample. The \texttt{Launch Clock} first must be converted to a logic signal, which is performed by the \texttt{Transition Generator}.

\subsubsection*{Transition Generator:}
This component is responsible for sending positive (0 $\rightarrow$ 1) and negative (1 $\rightarrow$ 0) transitions through the the \texttt{Route Under Test}, \texttt{Carry Chain}, and into the \texttt{Capture Registers}. The same $\theta$ defines the relationship between the signal traversing the \texttt{Route Under Test}/\texttt{Carry Chain} is launched and when it measured in the \texttt{Capture Registers}.
When the sensor is loaded onto an FPGA, $\theta$ is set to 0; an offset of $\theta$ is consistent between sensor design loadings.
When $\theta$ is set correctly, a transition will be propagating through the delay line when the \texttt{Capture Registers} are clocked and record a metastable transition region.
The distance that transition propagates is called the propagation distance, and is related to the propagation delay of the logic in the \texttt{Route Under Test} and \texttt{Carry Chain}.

\subsubsection*{Route Under Test:}
The primary intent of the TDC is to measure the timing delay through some FPGA programmable route that is affected by burn-in.
When $\theta$ is properly configured the output of the \texttt{Capture Registers} reflects how far a rising or falling signal has propagated, and consequently the timing delay though the \texttt{Route Under Test}.
BTI degradation causes the propagation delay to increase. The propagation delay decreases during BTI recovery.

\subsubsection*{Carry Chain:}
The primary structure of the TDC is a long linear array of delay elements. This is formed by a series of combinatorial logic elements that are able to propagate rising (0 $\rightarrow$ 1) and falling transitions (1 $\rightarrow$ 0). Ideally, each element is identical, with a timing delay of $\tau$, so that the propagation of signals is uniform at every stage of the \texttt{Carry Chain}. To ensure consistency throughout the chain,  the delay elements are uniformly placed and routed in consecutive physical locations. Our chosen sensor uses the fast look-ahead CARRY primitives of the Xilinx FPGA devices to construct this chain.

\subsubsection*{Capture Registers:}
Each element of the \texttt{Carry Chain} is output to a register, forming the \texttt{Capture Registers}. These registers are activated synchronously by the \texttt{Capture Clock}, that performs a capture of the state of the \texttt{Carry Chain}. If a rising (0 $\rightarrow$ 1) or falling transition (1 $\rightarrow$ 0) is propagating through the \texttt{Carry Chain}, and the \texttt{Capture Registers} are activated, the distance that signal has traveled will be captured. 

\subsubsection*{Propagation Distance:}
Each falling and rising transition is captured at the output registers, as shown at the bottom of Figure~\ref{fig/sensor_architecture}. Rising Transition 0 shows that the 0 $\rightarrow$ 1 transition reached \texttt{Output[38]}. Falling Transition 0 shows that the 1 $\rightarrow$ 0 propagated to somewhere between \texttt{Output[21]} and \texttt{Output[23]}, with some metastability between the two points. Rising Transition 1 propagates differently; the 0 $\rightarrow$ 1 transition propagates to between \texttt{Output[36]} and \texttt{Output[39]}. Similarly, Falling Transition 1 propagates to between \texttt{Output[20]} and \texttt{Output[23]}. These changes can represent deviations in the timing through the \texttt{Route Under Test} (i.e. burn-in). 

\subsubsection*{Post Processing:}
The output of the \texttt{Capture Registers} can be processed into a single value that represents the propagation time of a signal through a \texttt{Route Under Test}. This is done by computing the \textit{Binary Hamming Distance} of the output registers. This is defined for the rising transitions as the binary Hamming distance from \texttt{64'h\_0000\_0000\_0000\_0000}, and for falling transitions, the binary Hamming distance from \texttt{64'h\_ffff\_ffff\_ffff\_ffff}. The \textit{Binary Hamming Distance} of the example samples in Figure~\ref{section/sensor_structure} will yield the sequence: 39, 22, 38, 22.
\section{Experimental Setup}
\label{section/experimental_design}
We perform a series of experiments to determine the extent of BTI effects on programmable routing and the ability to recover the digital data that was previously held on those routes. We perform the first experiments locally on a ZCU102 Ultrascale+ board. Subsequent experiments are performed remotely on AWS F1 instances.  
The experiments use a target design and a measurement design described in Section~\ref{section/experimental_design/architectures}.  Section~\ref{section/experimental_design/phases} details the three experimental phases that calibrate the sensor, conditions (performs burn-in) on some target routes, and measures the BTI effects on the target routes. Section~\ref{subsection/open_titan} determine the approximate target route lengths for assets from the OpenTitan hardware root of trust. The experiments in Section~\ref{section/experimental_results} demonstrate the ability to execute Threat Models 1 and 2 remotely on an AWS F1 instance. 

\subsection{Experimental Designs}
\label{section/experimental_design/architectures}
Our experiments use two independent FPGA designs. The \burnindesign design holds the routes under test at a constant 1 or 0 value for a pre-determined duration to induce BTI effects. The \textit{Measure} design records changes in route propagation delays caused by BTI effects. The data held on these routes represent the Type A or B logical data that an attacker wishes to recover.

\begin{figure}[!htb]
\begin{subfigure}[t]{\linewidth}
  \centering
  \includegraphics[width=\linewidth]{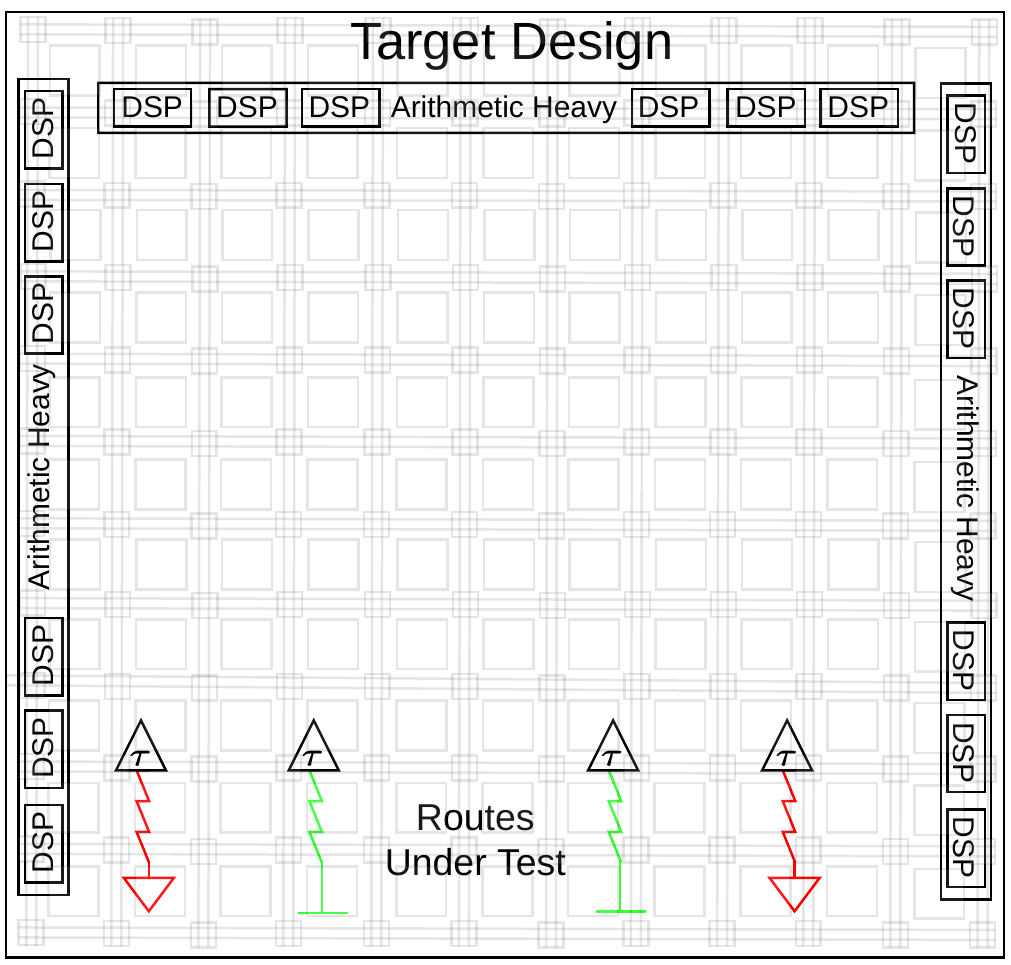}
\end{subfigure}
\caption{The \textbf{Target Design} conditions a set of pre-determined routes to 1/VCC (green) or 0/GND (red) aka the \textit{burn value}. This induces a BTI effects on transistors of each route. The \texttt{Arithmetic Heavy} circuit is used emulate the surrounding logic of many FPGA computations, and also increases on-chip temperature to accelerate BTI. The center of the design is left empty; it will be used in the measure design (Figure~\ref{fig/measure_architecture}).
}
\label{fig/burn-in_architecture}
\vspace{-15px}
\end{figure} 
\subsubsection*{Target Design:}
Figure~\ref{fig/burn-in_architecture} presents the \burnindesign design that biases a set of routes by holding them to a fixed 0 or 1 value.
We represent a logical 1 bias with the color green and logical 0 bias with the color red. 

The region of slices above the routes under test is explicitly left uninitialized (no logic may be placed there) during the compilation process. These slices will be used by the \textit{Measure} design for the placement of \texttt{Carry Chains}. Using these slices could introduce noise, or worse, erroneous results, into our propagation delay measurements of the test route. In theory this noise would be minimal as the length of the route through these slices is significantly shorter than the route being tested for BTI. While this does not negate all possible burn-in effects in the slices, since the vendor-determined state of an uninitialized slice could also introduce BTI effects, it at least suggests that all slices will be affected equally. This is consistent with \textbf{Threat Model 1}, as the attacker may be the one publishing the maliciously constructed AFI, and thus has the control to leave these slices empty. 

The routes containing \textbf{Type A} and \textbf{Type B} information is surrounded by other computation. 
We generalize these structures as \texttt{Arithmetic Heavy} circuits implemented as arrays of logic performing a pipelined fused multiply-add operation (similar to a machine learning or lattice cryptography accelerator).
This has the added benefit of accelerating the BTI effect through increased heat generation.

\begin{figure}[!htb]
\begin{subfigure}[t]{\linewidth}
  \centering
  \includegraphics[width=\linewidth]{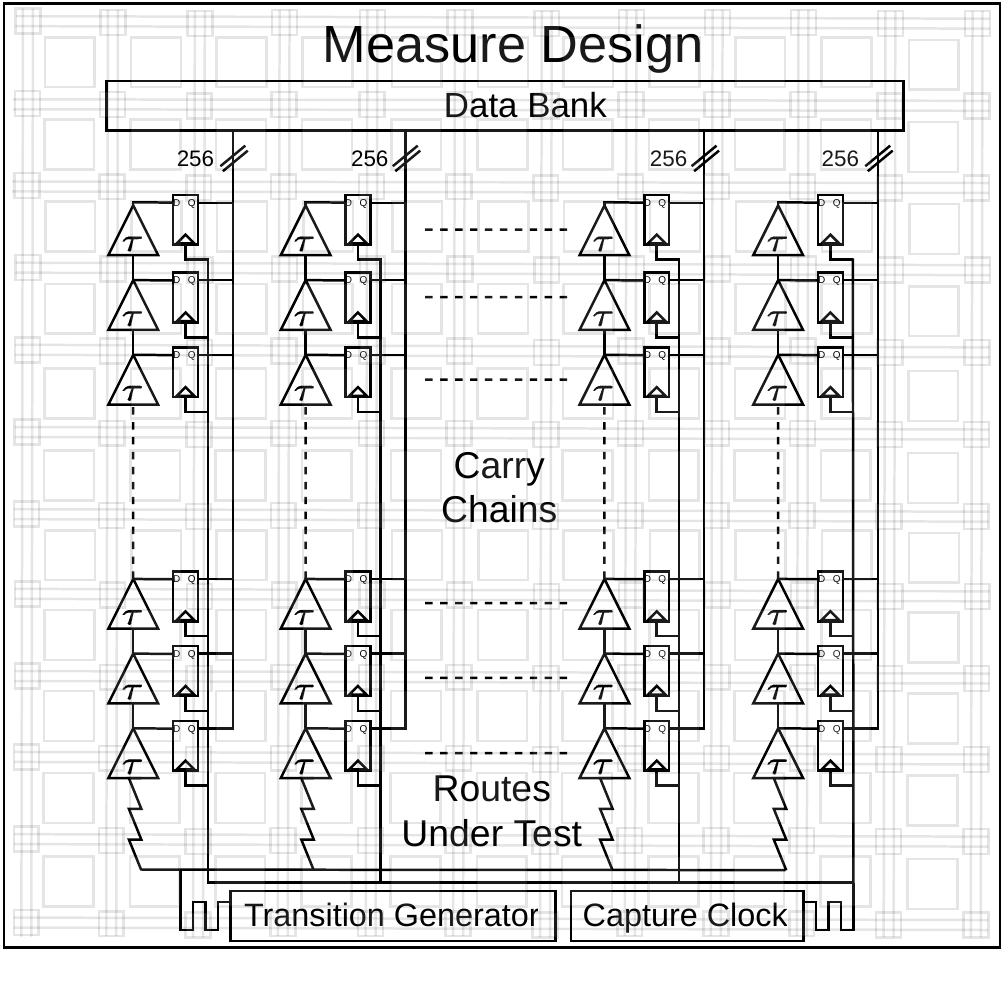}
\end{subfigure}
\caption{The \textbf{Measure Design} records the BTI degradation of multiple \texttt{Routes Under Test} using TDC sensors. As per Section~\ref{section/sensor_structure}, the \texttt{Transition Generator} is used to send rising ($0\rightarrow1$) and falling ($1\rightarrow0$) transitions through the tested routes. The changing propagation delay of these signals indicates the BTI effects on that route. 
}
\label{fig/measure_architecture}
\vspace{-15px}
\end{figure} 
\subsubsection*{Measure Design:}
Figure~\ref{fig/measure_architecture} presents a high level view of the architecture for measuring the propagation delay of the routes under test. These routes represent the \textbf{Type A} or \textbf{Type B} data an attacker intends to recover. By tracking the change in this propagation delay caused by BTI degradation and recovery, the side channel can be exposed to exploit \textbf{Threat Model 1} and \textbf{2}.

Each of these routes is a single \texttt{Route Under Test} from Figure~\ref{fig/sensor_architecture} and we instantiate an array of TDCs as presented in Section~\ref{section/sensor_structure}. No extraneous routing outside of the the \texttt{Route Under Test} is used to connect the \texttt{Transition Generator} to the \texttt{Carry Chain}.  Identical routing constraints from the \burnindesign design are used to generate the routes for the \textit{Measure} design. The routes test a variety of lengths and placements in order to build a general understanding of how burn-in is affected by route characteristics.

\subsection{Experimental Phases}
\label{section/experimental_design/phases}
These designs will then be used to form three experimental phases: \textit{Calibration} to configure the TDCs of the \textit{Measure} design, \textit{Condition} to induce the BTI effect on a predefined set of routes, \textit{Measurement} to measure BTI effects. 

\subsubsection*{Calibration Phase:} 
\textit{Calibration} is the first phase, and determines a baseline $\theta$ value that captures rising and falling transitions traveling through the tested routes into the \texttt{Carry Chain} and \texttt{Capture Registers}.
The TDC alone does not provide an absolute measure of the change in propagation delay through a tested route.
We can use the TDC determine the change in the propagation delay by examining the increase or decrease over time of the {\it Binary Hamming Distance} output of the sensor tuned to this baseline $\theta$ value. 
To create a baseline, a short series of $2^4$ samples (called a trace) is taken from each TDC as $\theta$ is iteratively reduced until the rising and falling transition appear in the output registers. 
We call this value $\theta_{init}$, and an individual value is computed and saved for every route under test in the \textit{Measure} design.

\begin{table*}[!ht]
    \centering
 
    \caption{\textbf{OpenTitan Earl Grey Distribution of Route Lengths (ps) on a Virtex UltraScale+.} This table reports the distribution of route lengths (in ps) for a selection of twenty security-critical assets in an OpenTitan Earl Grey implemented on a Virtex UltraScale+. Assets are sorted in ascending order by maximum route length. Route lengths of more than 1000 ps are common, and would increase when OpenTitan shares an FPGA with other logic.}
    {
    \begin{tabularx}{\textwidth}{|c|X|c|c|c|c|c|c|c|c|c|}
    \toprule
        \textbf{\#} & \textbf{Asset Paths} & \textbf{Type} & \textbf{Bus Width} & \textbf{MEAN} & \textbf{SD} & \textbf{MIN} & \textbf{25\%} & \textbf{50\%} & \textbf{75\%} & \textbf{MAX} \\
    \midrule
         1 & /otp\_ctrl\_otp\_lc\_data[state]                          & SV/T & 320 &  169.5 &  98.1 &   39 &   95.5 & 157.5 & 228 & 509 \\
         2 & /u\_otp\_ctrl/otp\_ctrl\_otp\_lc\_data[test\_exit\_token] & SV/T & 128 &  197.5 & 115.4 &   37 &  114   & 170 & 242.2 & 534 \\
         3 & /otp\_ctrl\_otp\_lc\_data[rma\_token]                     & SV/T & 101 &  239.8 & 122.8 &   38 &  148   & 222 & 325 & 583 \\
         4 & /otp\_ctrl\_otp\_lc\_data[test\_unlock\_token]            & SV/T & 128 &  207.9 & 120.1 &   38 &  130.5 & 178.5 & 247.2 & 609 \\
         5 & /keymgr\_aes\_key[key][1]\_282                            & CK   &  32 &  538.3 & 106.4 &  380 &  433.5 & 551 & 614 & 738 \\
         6 & /keymgr\_otbn\_key[key][0]\_285                           & CK   & 384 &  219.8 & 150.9 &   41 &   99   & 167 & 327.2 & 919 \\
         7 & /keymgr\_kmac\_key[key][0]\_28                            & CK   & 256 &  317.6 & 141.7 &   49 &  213.8 & 291 & 408 & 1050 \\
         8 & /otp\_ctrl\_otp\_keymgr\_key[key\_share0]                 & CK   & 256 &  187.3 & 200.8 &   37 &   54   & 109 & 217 & 1064 \\
         9 & /u\_otp\_ctrl/part\_scrmbl\_rsp\_data                     & CK   &  64 &  353.4 & 146.1 &  116 &  267.2 & 348.5 & 411.2 & 1075 \\
        10 & /keymgr\_aes\_key[key][0]\_283                            & CK   & 256 &  360.3 & 154.2 &   86 &  270   & 333 & 412.2 & 1311 \\
        11 & /u\_otp\_ctrl/u\_otp\_ctrl\_scrmbl/gen\_anchor\_keys      & CK   & 135 &  220.1 & 358.7 &    0 &   57   & 94 & 162.5 & 1333 \\
        12 & /otp\_ctrl\_otp\_keymgr\_key[key\_share1]                 & CK   & 256 &  262.5 & 273.4 &   37 &   51   & 158 & 335.5 & 1381 \\
        13 & /csrng\_tl\_rsp[d\_data]                                  & S    &  32 & 1291.8 & 105.7 & 1031 & 1244.8 & 1323 & 1359.8 & 1432 \\
        14 & /aes\_tl\_rsp[d\_data]                                    & S    &  32 & 1105.3 & 411.4 &  276 & 1135.8 & 1279 & 1369.5 & 1631 \\
        15 & /keymgr\_otbn\_key[key][1]\_284                           & CK   &  32 & 1062.7 & 281.2 &  480 &  854   & 1074.5 & 1270 & 1670 \\
        16 & /u\_otp\_ctrl/part\_otp\_rdata                            & S    &  64 & 1298.9 & 213   &  933 & 1118.5 & 1311.5 & 1447.2 & 1784 \\
        17 & /flash\_ctrl\_otp\_rsp[key]                               & CK   & 128 & 1816.6 & 404.6 & 1215 & 1503   & 1717.5 & 2010.2 & 3245 \\
        18 & /kmac\_app\_rsp                                           & S    & 777 &   94.2 & 179.7 &   15 &   40   & 58 & 97 & 3398 \\
        19 & /flash\_ctrl\_otp\_rsp[rand\_key]                         & CK   & 128 & 1908.1 & 670.7 &  553 & 1337   & 1882 & 2308.8 & 3706 \\
        20 & /aes\_tl\_req[a\_data]                                    & S    &  32 & 2114.8 & 471.8 & 1455 & 1805   & 2079.5 & 2337.2 & 3946 \\
    \bottomrule
    \end{tabularx}
    }
    \label{table/ot_hist_table_redux}
\end{table*}

\subsubsection*{Condition Phase:}
The \burnindesign design is loaded onto the FPGA and a pre-defined set of \textit{burn values} is applied to the routes under test.
These \textit{burn values} will either be \textbf{Type A} or \textbf{Type B} information that will induce variable BTI effect based on their value.
The \texttt{Arithmetic Heavy} circuits are activated in this phase to emulate user computation and exacerbate BTI degradation.

\subsubsection*{Measurement Phase:}
The \textit{Measurement} phase loads the \textit{Measure} design and tunes all of the TDCs to their respective $\theta_{init}$. Ten traces are taken from each TDC as $\theta$ is iteratively decreased from $\theta_{init}$, to avoid relying on a single trace that could be affected by architectural irregularities~\cite{favi200917ps,glamovcanin2020cloud, drewes2023turnon}. 
For each route, the mean {\it Binary Hamming Distance} is computed on across all samples within each trace, and then the mean of all traces is computed to obtain a single value representing the propagation delay through the route under test.
This value is converted to a measure of time based on a derived relationship of $\frac{2.8ps}{bit}$ for UltraScale+ parts~\cite{drewes2023turnon, xilinx_ultrascale_clocking}.
Any deviation in this value represents BTI-induced variation on a route.

\subsection{OpenTitan Hardware Root of Trust}
\label{subsection/open_titan}

We study the OpenTitan hardware root of trust (RoT) to provide a context for a realistic target for our threat models. OpenTitan is a commercial-grade, open-source hardware root of trust~\cite{OpenTitan}. The OpenTitan Earl Grey  integrated into systems to carry out core, security-critical functionalities related to trusted platform module (TPM), platform integrity, and 2nd factor authentication. 
The OpenTitan consists of a security-enhanced RV32IMCB RISC-V Ibex core, cryptographic IP cores (e.g., AES, KMAC, HMAC), and memories (e.g., ROM, eFLASH, SRAM, OTP) protected by access control mechanisms.

OpenTitan is an open-source hardware design; all design files are available online. OpenTitan encourages a design flow where the user solely modifies the boot ROM (the data used to initialize the FPGA memory) of a pre-built bitstream. Thus, it adheres to threat model assumption 1 (Section~\ref{section/threat_model}) that the assets locations are known.  

OpenTitan has many important security assets that govern its operation. Assets include cryptographic keys to encrypt data stored in off-chip memories (e.g., one-time programmable memory), keys to scramble data before transmission across on-chip busses to limit power side channels, and life-cycle related state values/tokens for attestation, identity management, and debug control. 

We identify twenty security-critical assets in the following groups:
\begin{itemize}
    \item \textbf{Cryptographic Keys (CK):} OpenTitan has a number of cryptographic keys that need to be protected that are spread across the design. This includes keys stored in the one-time programmable (OTP) memory and OpenTitan Key Manager. The OTP controller has an access control mechanism that arbitrates OTP data accesses and buffers key values;
    \item \textbf{State Values or Tokens (SV/T):} Assets stored in one-time-programmable (OTP) memory for use in the life-cycle controller, which influences the OpenTitan's DFT functionality, NVM backdoor access, debug, and CPU functionality; 
    \item \textbf{Signals (S):} Variables carrying sensitive information from/to security peripherals.
\end{itemize}

Table~\ref{table/ot_hist_table_redux} reports the route length distribution of twenty security-critical assets in OpenTitan implemented on a Virtex UltraScale+. \textbf{Bus Width} records the number of routes associated with each asset. \textbf{MEAN} and \textbf{SD} are the mean route length and standard deviation for each asset's routes, respectively. \textbf{MIN} is the minimum route length for each asset. \textbf{25\%}, \textbf{50\%}, and \textbf{75\%} are the route length for the 25th, 50th, and 75th percentiles of each asset's routes, respectively. \textbf{MAX} records the maximum route length for each asset. The assets are sorted in ascending order according to \textbf{MAX} route length.

Most routes are short -- only a few hundred picoseconds. However, there are longer route lengths that approach 4 ns.
When integrated with other cores or accelerators, it is logical that these route lengths will increase. 
\section{Experimental Results}
\label{section/experimental_results}

The experiments interleave the calibration, condition, and measurement phases (Section~\ref{section/experimental_design/phases}) to extract pentimenti -- BTI-induced side channels in FPGA programmable routing.
The target route lengths are informed by a study of  OpenTitan asset delay in Section~\ref{subsection/open_titan}.
Experiment 1 uses a new ZCU102 Ultrascale+ FPGA development board.
This allows us to characterize the burn-in effect while controlling temperature, eliminate previous FPGA usage, and minimize system computation during measurement. The remote cloud platform provides substantially less control over environmental conditions.
Experiment 2 validates \textbf{Threat Model 1} on the AWS F1 platform.
Experiment 3 validates \textbf{Threat Model 2} on the AWS F1 platform.

\subsection{Experiment 1 (Lab Environment)}
\label{section/experimental_results/zcu102}
Experiment 1 studies BTI degradation and recovery effects on a local, new FPGA. The experiment validates that the burn-in degradation occurs and is differentiable, making \textbf{Threat Model 1} possible. Additionally, it shows that BTI is non-permanent. BTI recovery is observable and measurable, which is required for \textbf{Threat Model 2}.

A new ZCU102 Ultrascale+ is placed in a temperature-controlled forced convection oven (Lab Companion OF-01E) set to 60$^\circ$C. The oven maintains a constant temperature, which ensures that temperature changes do not influence the delays. The ZCU102 is factory new; it will experience the largest BTI effects since no degradation has occurred.

64 routes are studied on the ZCU102 Ultrascale+ development board. The first group of 16 routes each has a delay of 1000 ps, the second 2000 ps, the third 5000 ps, and the fourth 10000 ps delay. 
The delay relates to the number of transistors used to form that programmable route subject to BTI effects. 

Experiment 1 starts with a 200-hour burn-in period of the 64 routes. A randomly generated value $X$ is applied to the routes. $X$ is held constant over the entire 200 hours. The goal is to induce burn-in on the routes and understand the extent of BTI effects. Burn-in is followed by a 200-hour recovery period that applies a constant $\overline{X}$ to the routes under test to induce BTI recovery. 

Experiment 1 is divided into three experimental periods consisting of phases from Section~\ref{section/experimental_design/phases}:
\begin{itemize}

\item \textbf{Hour 0:}
The \textit{Calibration} phase is executed to compute the $\theta_{init}$ for each of the 64 routes.

\item \textbf{Hours [0,200):}
The burn-in period alternates between \textit{Condition} and \textit{Measurement} phases. The \textit{Condition} phase applies the \textit{burn values} $X$ to the 64 routes for one hour.
Then, the \textit{Measurement} phase is launched, which tunes the phase difference between the transition generator and capture clock ($\theta_{init}$) to ensure the transition falls within the carry chain. The TDC sensors capture data for each of the 64 routes under test as described in Section~\ref{section/experimental_design/architectures}.
This condition/measurement sequence is repeated $200 \times$ (approximately $200$ hours). The \textit{Measurement} phase runs once per hour to record the BTI degradation effects. Measurement is fast, taking less than a minute. Thus the vast majority of the time is spent in the condition phase.

\item \textbf{Hours [200,400):}
The recovery period inverts the values previously held on the routes under test.
This is nearly identical to hours [0,200), except the \textit{Condition} phase loads $\overline{X}$, the complement of $X$, into the 64 routes.
This period focuses on understanding BTI recovery effects.
\end{itemize}

\begin{figure}[!htb]
\centering
\begin{subfigure}[t]{\linewidth}
    \centering
    \includegraphics[width=\linewidth]{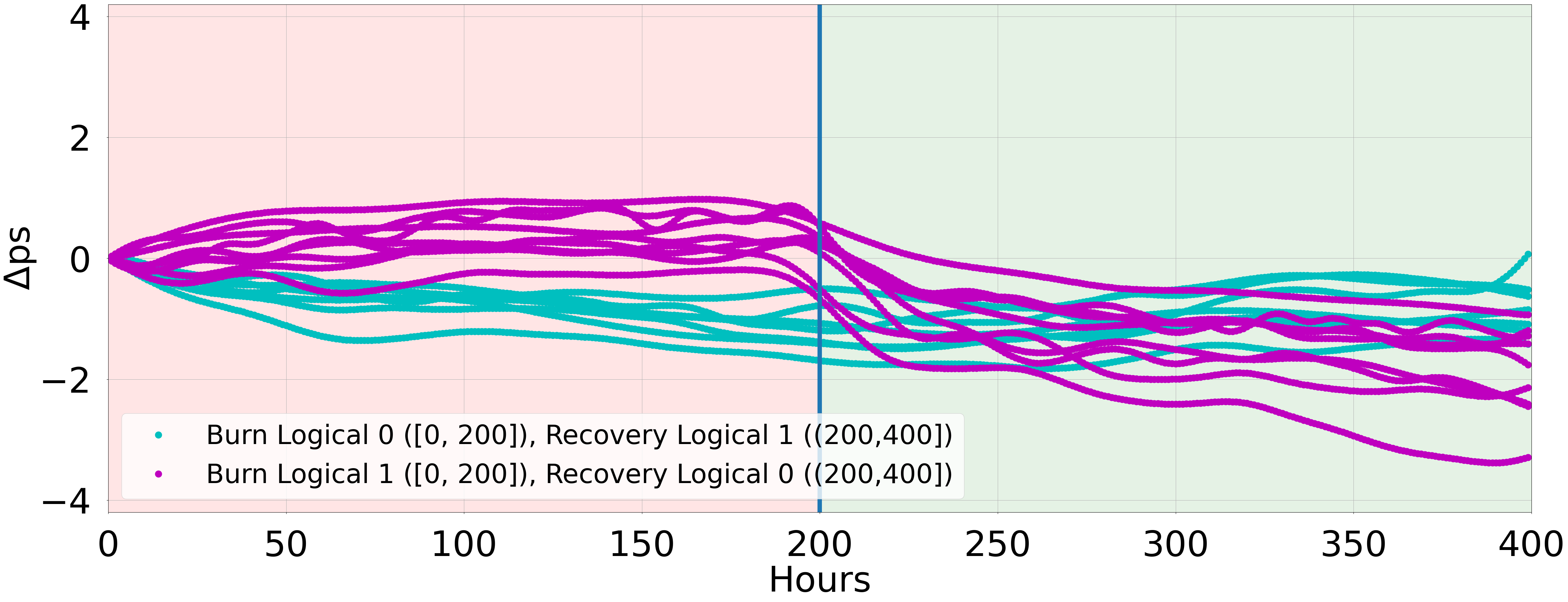}
    \vspace{-18px}
    \caption{1000ps Routes}
    \label{fig/zcu102_burnin_recover_0}
\end{subfigure}
\begin{subfigure}[t]{\linewidth}
    \centering
    \includegraphics[width=\linewidth]{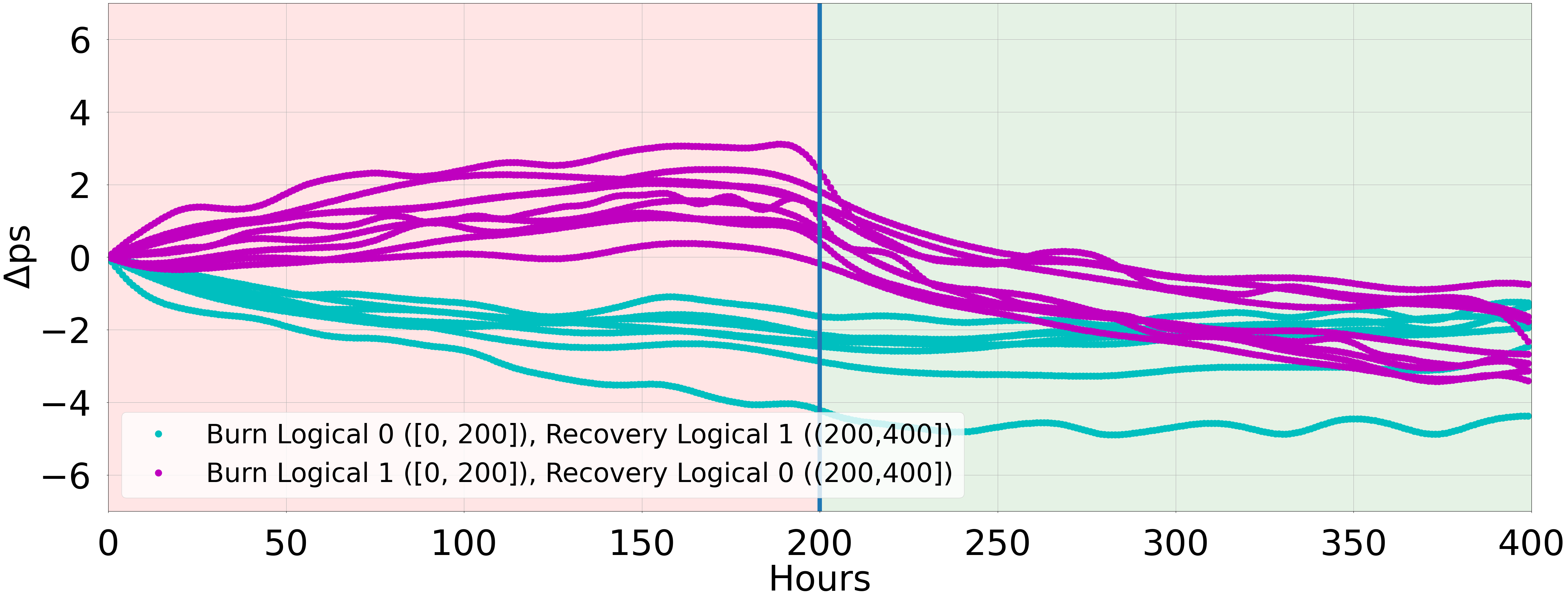}
    \vspace{-18px}
    \caption{2000ps Routes}
    \label{fig/zcu102_burnin_recover_1}
\end{subfigure}
\begin{subfigure}[t]{\linewidth}
    \centering
    \includegraphics[width=\linewidth]{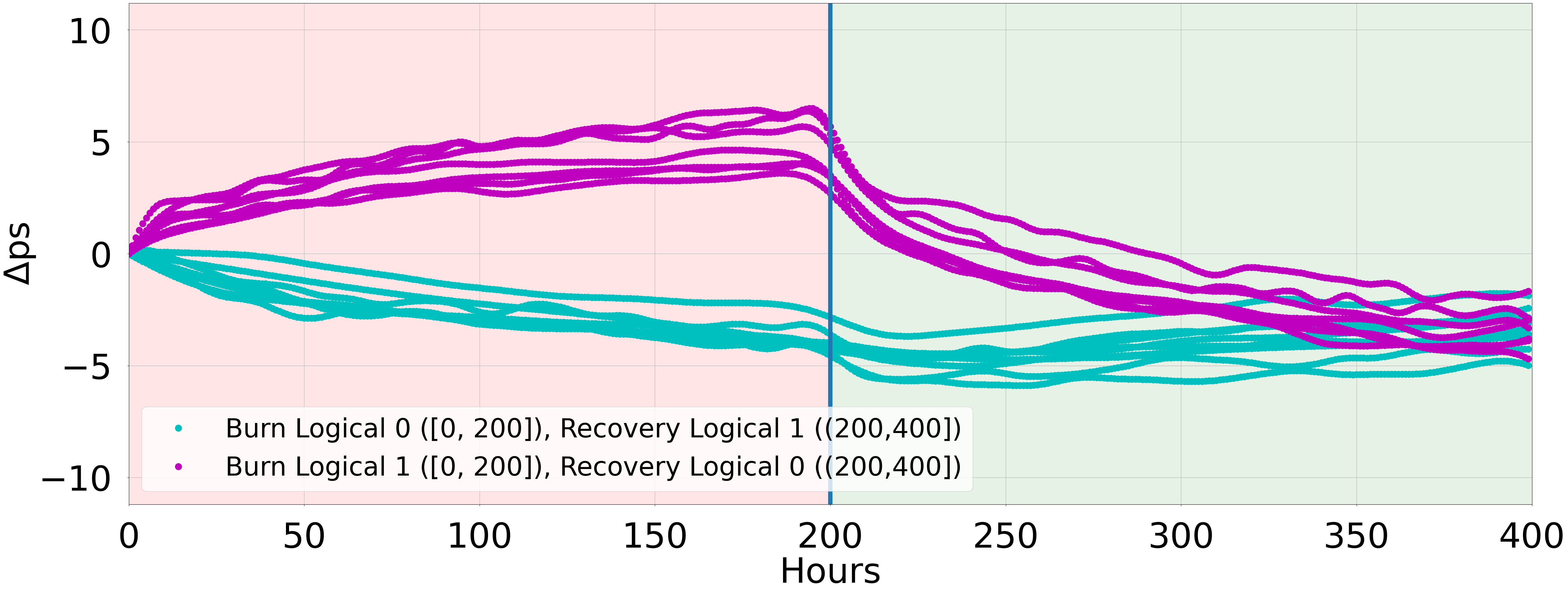}
    \vspace{-18px}
    \caption{5000ps Routes}
    \label{fig/zcu102_burnin_recover_2}
\end{subfigure}
\begin{subfigure}[t]{\linewidth}
    \centering
    \includegraphics[width=\linewidth]{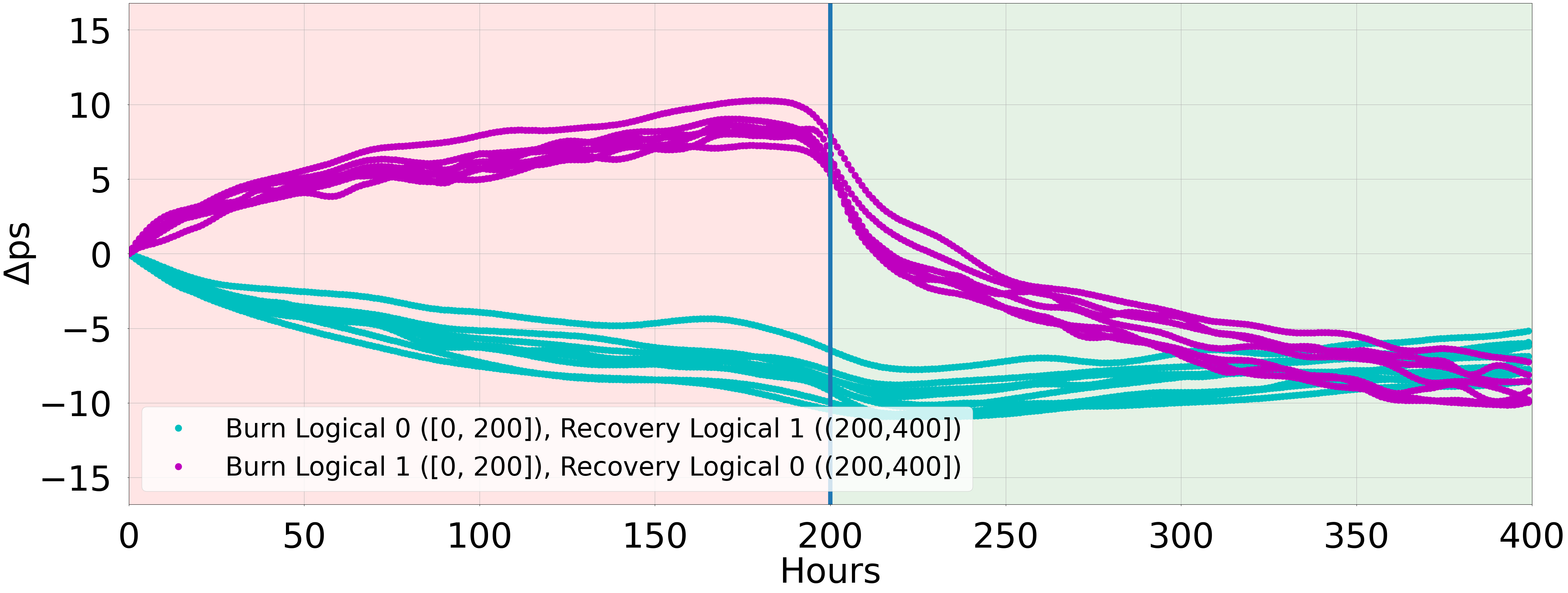}
    \vspace{-18px}
    \caption{10000ps Routes}
    \label{fig/zcu102_burnin_recover_3}
\end{subfigure}
\vspace{-5px}
\caption{\textbf{Experiment 1 (Lab Environment):} Sets of 16 routes of varying lengths are initialized on a new ZCU102 FPGA. The experiment occurs at 60$^\circ$C in a temperature-controlled environment. During the initial 200-hour burn-in period, a random burn value $X$ is conditioned into the routes (red background). After which, 200-hour BTI recovery is induced by applying $\overline{X}$ into the routes under test (green background). The timing difference between each route's falling and rising transition delay is recorded every hour. Routes conditioned with logical 0 behave differently than routes conditioned with logical 1 in the burn-in and recovery periods. This reveals the unique effects of BTI degradation and recovery. It indicates that \textbf{Threat Model 1} and \textbf{Threat Model 2} are possible. 
}


\label{fig/zcu102_burnin_recover}
\vspace{-15px}
\end{figure}

Figure~\ref{fig/zcu102_burnin_recover} plots the 400-hour results of Experiment 1 for four different route delays. The four graphs each have 16 routes with various delays -- (a) 1000ps, (b) 2000ps, (c) 5000ps, and (d) 10000ps. 
A switch from burn-in values $\overline{X}$ to recovery values $\overline{X}$ happens at the 200-hour mark, denoted by the transition between red and green backgrounds.

Every net is measured once per hour. A \textit{Measure} phase is loaded, and the TDC sensor readings are recorded. 
Ten traces are taken from each TDC as the phase shift $\theta$ is decreased to avoid architectural irregularities~\cite{favi200917ps, glamovcanin2020cloud, drewes2023turnon}. 
We record the rising and falling transitions within each trace.
For each transition type, the mean propagation distance is computed across all samples within a trace.
Then, we compute the mean of the ten traces to obtain a single value representing the propagation delay through the route under test.
Next, we subtract the rising transition distance from the falling transition distance to isolate the BTI effect to the \emph{Route Under Test}.
This bit value is converted to seconds based on a derived relationship of $\frac{2.8ps}{bit}$ for UltraScale+ parts~\cite{xilinx_ultrascale_clocking, drewes2023turnon}.
Finally, we center the data to the point at hour zero; any deviation from zero represents BTI degradation or recovery-induced variation on a route, which we call $\Delta$ ps. Measurement took about 52 seconds for all the routes in our cloud experiments; only $1.4\%$ of the overall time is spent performing measurements.

The choice to plot the falling minus rising is based on the fact that architecturally the falling (rising) transition stresses the NMOS (PMOS) transistors in a route.
This is due to NMOS (PMOS) devices being best suited to passing 0 (1) values.
And so, the timing difference between the rising and falling signal reduces to a single value for each hour that considers both the NBTI and PBTI degradation. Figure~\ref{fig/bti_high_level} discusses this in more depth.

Datapoints are colored cyan if their burn value $X$ is a logical 0 and purple if their burn value $X$ is a logical 1.
A red background indicates that the value applied to the routes is the burn value $X$, and green background is the BTI recovery period where the values are complemented $\overline{X}$.
The resulting time series are smoothed with a kernel regression, which finds a non-linear relationship between a pair of variables. Specifically, the Python statsmodels package's nonparametric kernel regression class is used in continuous mode with a local linear estimator.

A trend is immediately apparent in the first 200 hours (red half) of the charts in Figure~\ref{fig/zcu102_burnin_recover}.
Burn value 0 (cyan) routes decrease from hour zero.
Burn value 1 (magenta) routes increase from hour zero.
This trend occurs regardless of the length of the route, but the magnitude differs. 
While a 10000 ps route is on the longer end for most designs, 1000 ps routes are commonplace, e.g., see the OpenTitan study in Section~\ref{subsection/open_titan}. Furthermore, this helps us understand the limits of our strategy. 
There appear to be no limitations in route length as to observable burn-in effects, with the 1000 ps tested routes showing a clear difference between GND and VCC burned routes.

\emph{These results indicate that \textbf{Threat Model 1} is possible. If an attacker can observe BTI effects on a route before and after a design, they can easily deduce the burn value on a route and observe a side channel.}

Figure~\ref{fig/zcu102_burnin_recover_0} shows that the 1000 ps routes have $\pm [1,2]$ ps difference between the rising and falling transition at the 200-hour mark. 
The 2000 ps routes have a $\pm [2,3]$ ps difference (Figure~\ref{fig/zcu102_burnin_recover_1}), the 5000 ps routes have a $\pm [5,6]$ ps difference (Figure~\ref{fig/zcu102_burnin_recover_2}), and the 10000 ps routes have a $\pm [10,11]$ ps difference.
This data matches our expectation in burn-in behavior: the number of transistors in the route (i.e., the route length) directly increases the route’s delay.

At the 200-hour mark, the experiment moves from the burn-in to recovery. The condition route values change from $X$ to $\overline{X}$.
The routes with logical 1 burn-in in $X$ in the first 200 hours (and logical 0 in the recovery period) quickly return to their pre-burn state across all route lengths.
This recovery takes approximately 30-50 hours before the propagation delay difference between the rising and falling transition has returned to the original state at hour 0.
We do not see the same behavior in the routes that were logical 0 in the first 200 hours and logical 1 in the second 200 hours; they recover, but the process takes much longer (over 200 hours).

These results indicate that BTI is elastic and non-permanent.
In addition, we can see that the BTI recovery in routes conditioned by burn 1 is substantially faster than in routes with burn 0.
This pattern persists for all tested route lengths, suggesting a fundamental difference between the NBTI and PBTI effect on the 16nm FinFET transistors of the UltraScale+ device.
The difference in BTI recovery enables \textbf{Threat Model 2}.

The quick recovery of the burn 1 routes indicates that they might be easier to detect when targeting Type B data (Threat Model 2). 
A Threat Model 2 attacker obtains the FPGA during the recovery period.
They will not know the initial values and thus cannot complement them. 
The attacker must set the target values to logical 0 or 1.
Since the Burn 1 degradation values see the greatest and fastest recovery, the attacker would set all recovery values to condition to logical 0 to observe the quick recovery. 
This motivates us to set the recovery values to logical 0 in Experiment 3 (Section~\ref{section/experimental_results/f1_experiment_1}).

\subsection{Experiment 2 (Cloud Environment)}
\label{section/experimental_results/f1_experiment_0}

Experiment 2 tests the viability of \textbf{Threat Model 1} on an AWS F1 cloud FPGA, which aims to extract sensitive data from a rented third-party design. 
The attacker can load and unload the design and wants to extract design intellectual property, e.g., netlist constant holding cryptographic keys or neural network weights.  
The cloud environment provides no control over temperature, and it is likely the device is years old, making BTI effects less observable~\cite{alam2005comprehensive}. 
This experiment is performed in the eu-west-2 AWS region, which puts potentially fours years of wear on the device.\footnote{https://aws.amazon.com/about-aws/whats-new/2018/10/amazon-ec2-f1-instance-expands-to-more-regions-adds-new-features-and-improves-development-tools/}

We examine 16 1000 ps, 16 2000ps, 16 5000 ps routes and 16 10000 ps delay routes.
64 random bit values $X$ are applied to these routes in the condition phase. 
The \burnindesign and \textit{Measure} designs are built around these routes according to Section~\ref{section/experimental_design/architectures}. The \burnindesign design is configured to utilize 3896 DSPs for this architecture. 
The \burnindesign design consumes 63 W of a maximum of 85 W imposed by AWS.

Experiment 2 is divided into two periods:
\begin{itemize}
\item \textbf{Hour 0:}
The \textit{Calibration} phase is executed to compute the $\theta_{init}$ for each of the 64 routes.

\item \textbf{Hours [0,200]:}
During each hour, we run a long \textit{Condition} phase and a short \textit{Measurement} phase. The \textit{Condition} phase applies the \textit{burn values} $X$ to the 64 routes under test.
The \textit{Measurement} phase tunes TDCs to $\theta_{init}$ and captures delay estimates for each route as described in Section~\ref{section/experimental_design/architectures}. 
After this brief measurement process (33 seconds), the data is saved.
This is repeated 200 times (over 200 hours) to study \textbf{Threat Model 1}.
\end{itemize}

\begin{figure}[!htb]
\centering
\begin{subfigure}[t]{\linewidth}
    \centering
    \includegraphics[width=\linewidth]{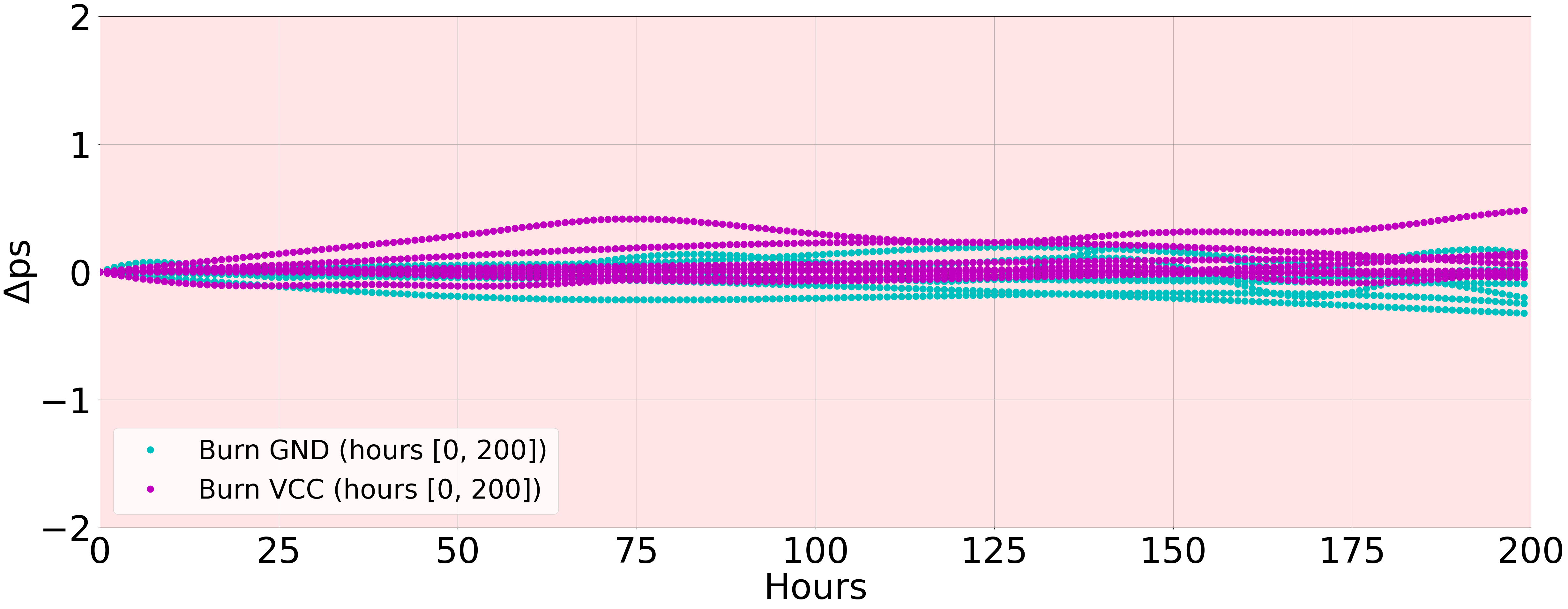}
    \vspace{-18px}
    \caption{1000 ps Routes}
    \label{fig/f1_experiment_1_2}
\end{subfigure}
\begin{subfigure}[t]{\linewidth}
    \centering
    \includegraphics[width=\linewidth]{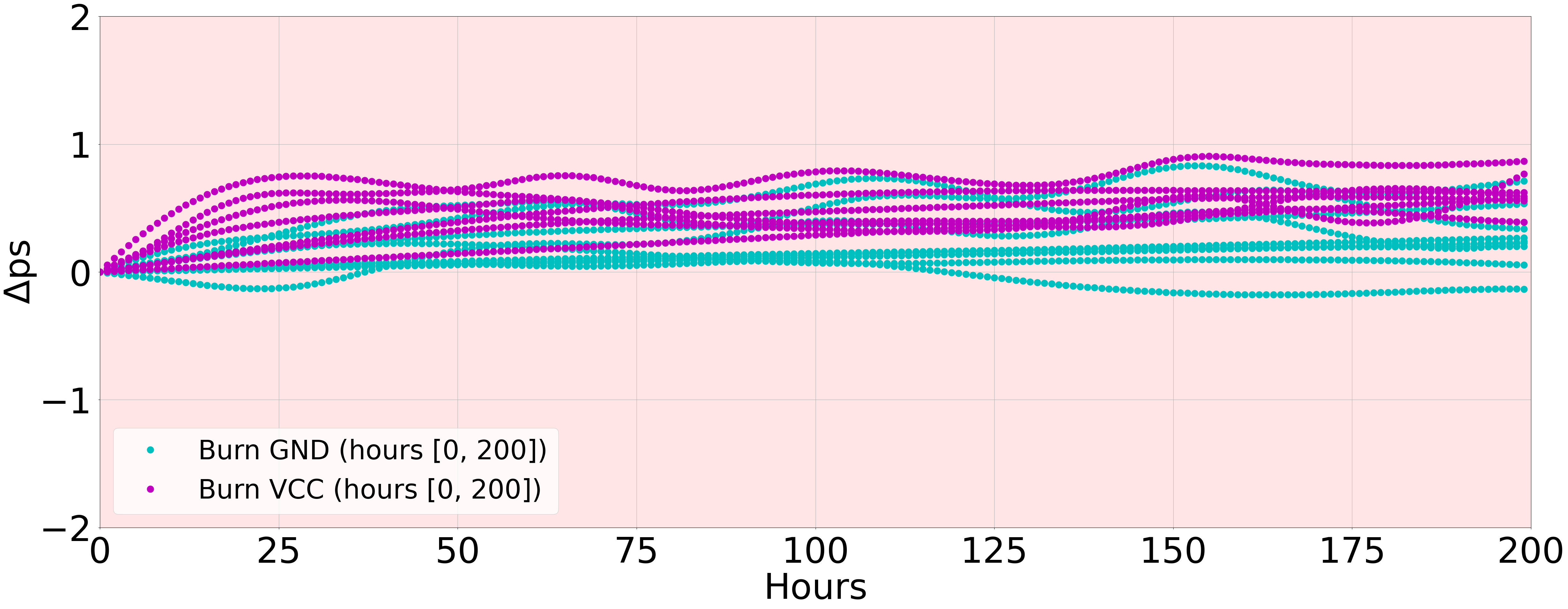}
    \vspace{-18px}
    \caption{2000 ps Routes}
    \label{fig/f1_experiment_1_3}
\end{subfigure}
\begin{subfigure}[t]{\linewidth}
    \centering
    \includegraphics[width=\linewidth]{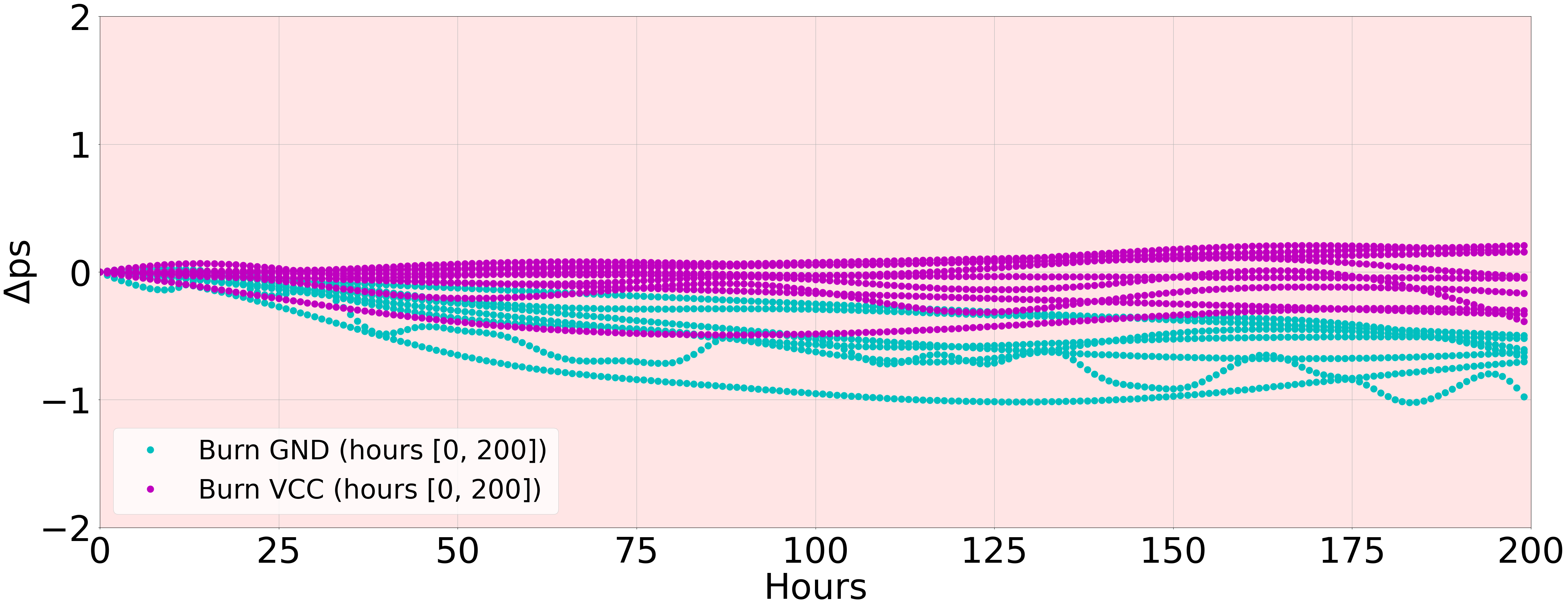}
    \vspace{-18px}
    \caption{5000 ps Routes}
    \label{fig/f1_experiment_1_0}
\end{subfigure}
\begin{subfigure}[t]{\linewidth}
    \centering
    \includegraphics[width=\linewidth]{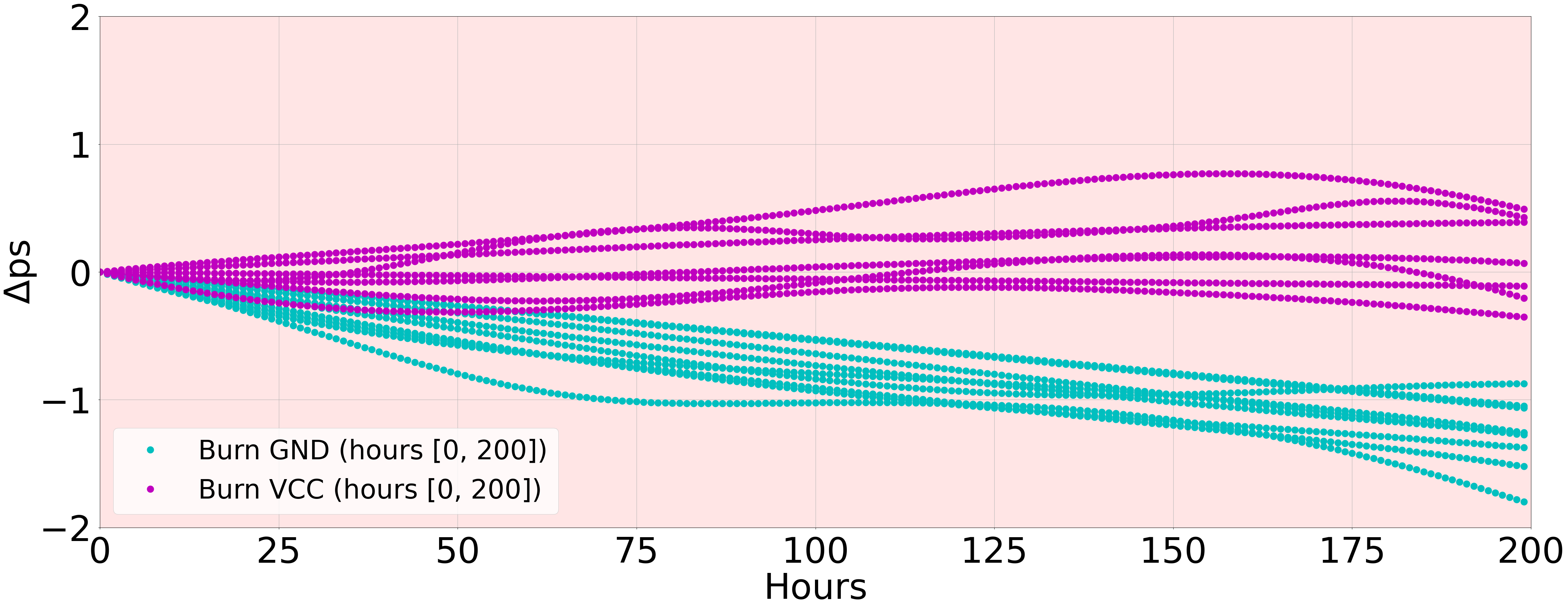}
    \vspace{-18px}
    \caption{10000 ps Routes}
    \label{fig/f1_experiment_1_1}
\end{subfigure}
\vspace{-5px}
\caption{
\textbf{Experiment 2 (Cloud Environment):} A random burn value $X$ is conditioned into four sets of FPGA routes. \ref{fig/f1_experiment_1_2} has 16 1000 ps routes, \ref{fig/f1_experiment_1_3} has 16 2000 ps routes, \ref{fig/f1_experiment_1_0} has 16 5000 ps routes and \ref{fig/f1_experiment_1_1} has 16 10000 ps routes. The propagation delay difference ($\Delta$ps) between the falling and rising transition is measured once per hour over 200 hours. Over time, the burn-in of the 0 (cyan) and 1 (purple) values produces observable changes in their route delays due to BTI degradation. This enables an attacker to recover \textbf{Type 1} data (Design Data) and execute \textbf{Threat Model 1.}
}
\label{fig/f1_experiment_1}
\vspace{-15px}
\end{figure}

Figure~\ref{fig/f1_experiment_1} shows the results from Experiment 2, testing \textbf{Threat Model 1} on the AWS F1 platform. 
$X$ is the \textbf{Type A} (Design Data) an attacker wishes to recover.
The values of $X$ are expressed in cyan for burn 0 values, and magenta for burn 1 values but are opaque to the attacker.

Figures~\ref{fig/f1_experiment_1_2}, \ref{fig/f1_experiment_1_3}, \ref{fig/f1_experiment_1_0}, \ref{fig/f1_experiment_1_1} demonstrate the same trends as in Experiment 1.
These results are expectedly noisier than from the ZCU102, which was a new part held at a constant temperature. 
The difference between the burn 0 rising and falling transitions decreases.
In contrast, the burn 1 routes behave the exact opposite; the propagation delay difference between the rising and falling transition increases.
This makes the routes easily distinguishable, with burn 0 (cyan) decreasing immediately from hour zero and burn 1 (magenta) increasing immediately from hour zero.
This pattern persists irrespective of the route length (1000ps vs. 2000ps vs. 5000 ps vs. 10000 ps), but the magnitude differs. 
By examining the trends in data, $X$ can be derived, demonstrating \textbf{Threat Model 1} recovery of \textbf{Type A} data is possible.

The 1000 ps routes of Figure~\ref{fig/f1_experiment_1_2} experience $\pm [0,.2]$ ps difference, the 2000 ps routes of Figure~\ref{fig/f1_experiment_1_3} experience $\pm [0,.4]$ ps difference, the 5000 ps routes of Figure~\ref{fig/f1_experiment_1_0} experience $\pm [0,1]$ ps difference, and the 10000 ps routes in Figure~\ref{fig/f1_experiment_1_1} experience $\pm [0,2]$ ps difference.
This further validates our previous conclusion that the amount of burn-in is proportional to the tested route length. It also indicates that the burn-in for the cloud FPGAs is lesser than that of the new ZCU102 from Experiment 1 (compare to Figure~\ref{fig/zcu102_burnin_recover_2} and \ref{fig/zcu102_burnin_recover_3}). This is not surprising, given that cloud FPGAs are older and more used. Thus, it is more challenging to extract pentimenti from cloud FPGAs than a local FPGA. 

It is now clear how \textbf{Threat Model 1} can be exploited by an attacker.
In this model, an attacker instantiates a design that contains \textbf{Type A} sensitive information on an AWS FPGA.
The attacker knows the location of sensitive data routes, and they construct a \textit{Measurement} design that maps TDCs to these routes.
The attacker can interleave measurements every hour to eventually exposes the \textbf{Type A} data based on the relationship between the rising and falling transition over time. The attacker can continue the burn-in process until they are satisfied that the sensitive values are extracted. 

\subsection{Experiment 3 (Cloud Environment)}
\label{section/experimental_results/f1_experiment_1}

Experiment 3 studies the viability of \textbf{Threat Model 2}, which exploits BTI recovery as a side channel. 
Sixteen 1000 ps, 2000 ps, 5000 ps, and 10000 ps delay routes are instantiated on an AWS F1 FPGA. The routes undergo a burn-in period during which we do not measure the time delay. Then, they switch to the recovery phase. 
We aim to understand if it is possible to recover pentimenti of a previous user of the FPGA by only measuring during the recovery period. This meets the assumptions required for Threat Model 2. $\theta_{init}$ is consistent across all FPGAs of the same type, and so capturing $\theta_{init}$ once on any board is sufficient to assume that $\theta_{init}$ is known a priori for any attack.  

Experiment 3 is divided into three periods:
\begin{itemize}

\item \textbf{Hours [0, 200):} 
A burn-in period induced by the victim computation. The \textit{Condition} phase executes with a constant, randomly generated $X$ value loaded into the 64 routes under test. Calibration is \textbf{not} performed. The attacker does not have control of the FPGA, and thus measurement is not allowed. The condition phase is run uninterrupted for 200 hours. 

\item \textbf{Hour 200:}
The victim relinquishes control of the FPGA, and the attacker gains control.

\item \textbf{Hours (200, 225]:} 
The attacker launches the \textit{Measurement} phase, which tunes TDCs to $\theta_{init}$, and captures traces of the 64 routes under test as described in Section~\ref{section/experimental_design/architectures}.
After this quick process, the data is saved.
Then, the attacker launches the \textit{Condition} phase that runs for one hour.
The Measurement/Condition sequence is repeated for 25 iterations (hours).
\end{itemize}

The target victim design holds a constant 64-bit $X$ value on FPGA routes for 200 hours without interruption. $X$ represents \textbf{Type B} (User Data) the attacker aims to recover. 
The burn-in $X$ values are opaque to the attacker. 
But, $X$ can be derived by the attacker based on the recovery behavior.

After 200 hours, the attacker gains control of the FPGA and thus can start measuring for BTI effects. The attacker sets all routes under test to logical 0 and measures the propagation delay once per hour over the next 25 hours. 
The attacker is looking for BTI recovery to extract pentimenti of the previously loaded design data.  
We do not assume the attacker has prior information about the FPGA before the victim computed upon it.

The choice of setting all routes to logical 0 is motivated by the results in Experiment 1; routes that were logical 1 in $X$ and were switched to $\overline{X}$ quickly returned to the original value. Thus, it exhibits a more significant signal for detection. An attacker could also choose to set all lines to logical 1 or a mixture of 0 and 1. 

\begin{figure}[!htb]
\centering
\begin{subfigure}[t]{\linewidth}
    \centering
    \includegraphics[width=\linewidth]{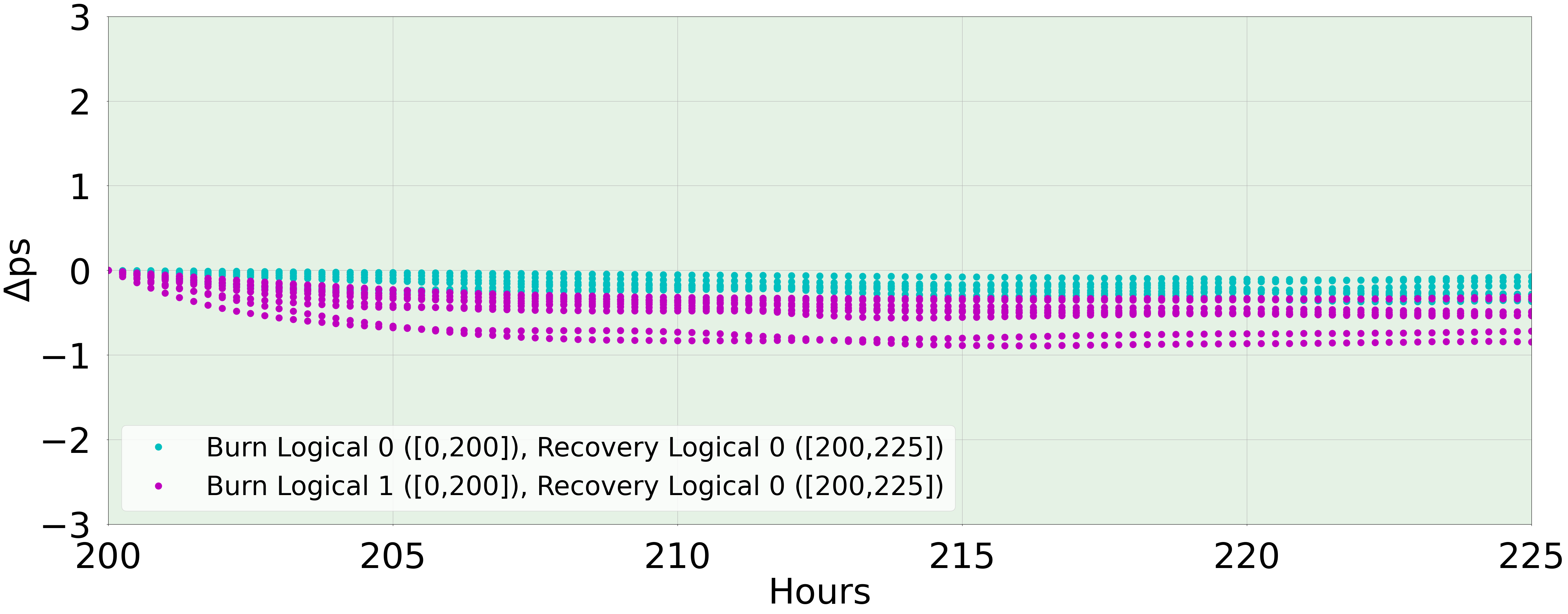}
    \vspace{-18px}
    \caption{1000 ps Routes}
    \label{fig/f1_experiment_2_2}
\end{subfigure}
\begin{subfigure}[t]{\linewidth}
    \centering
    \includegraphics[width=\linewidth]{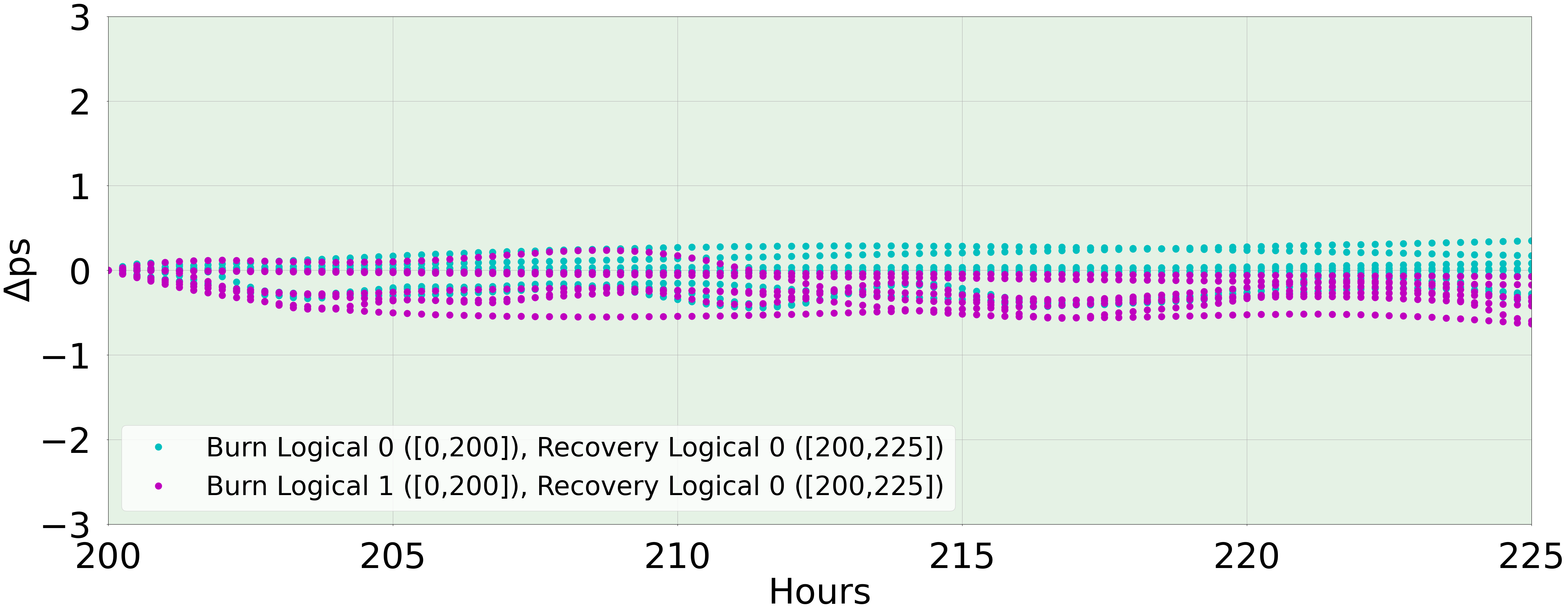}
    \vspace{-18px}
    \caption{2000 ps Routes}
    \label{fig/f1_experiment_2_3}
\end{subfigure}
\begin{subfigure}[t]{\linewidth}
    \centering
    \includegraphics[width=\linewidth]{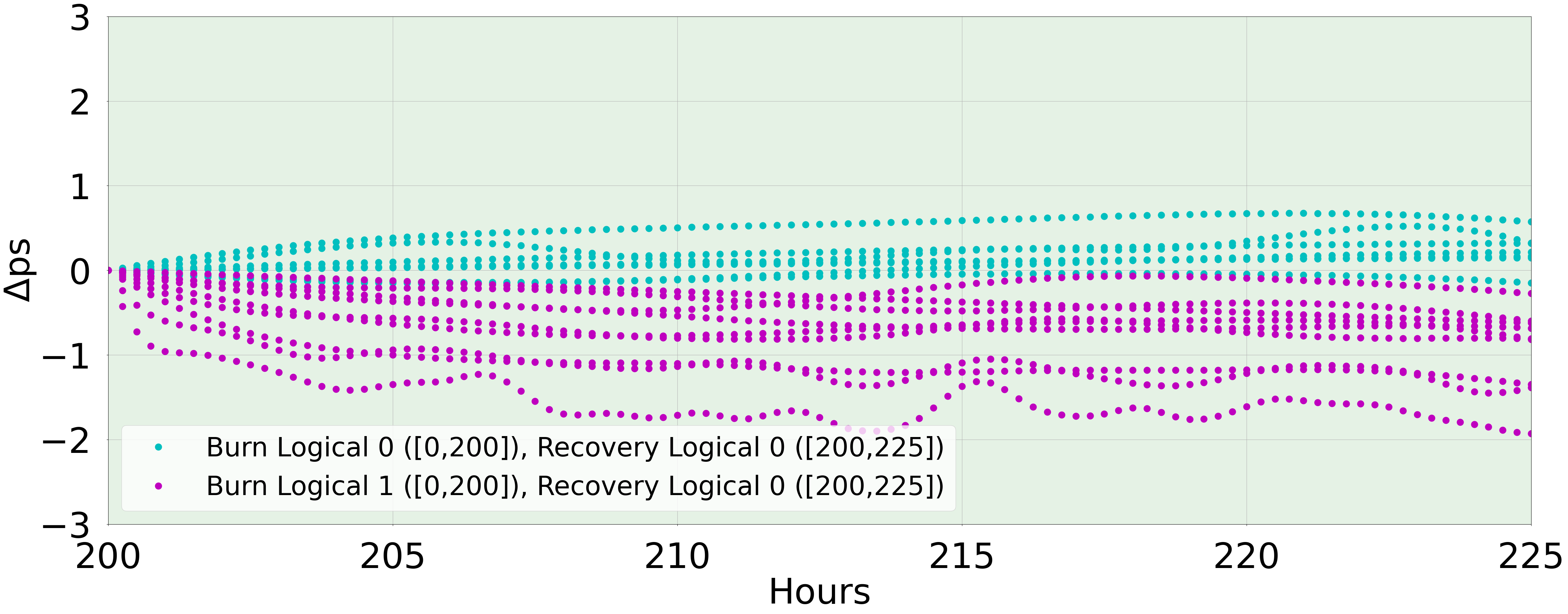}
    \vspace{-18px}
    \caption{5000 ps Routes}
    \label{fig/f1_experiment_2_0}
\end{subfigure}
\begin{subfigure}[t]{\linewidth}
    \centering
    \includegraphics[width=\linewidth]{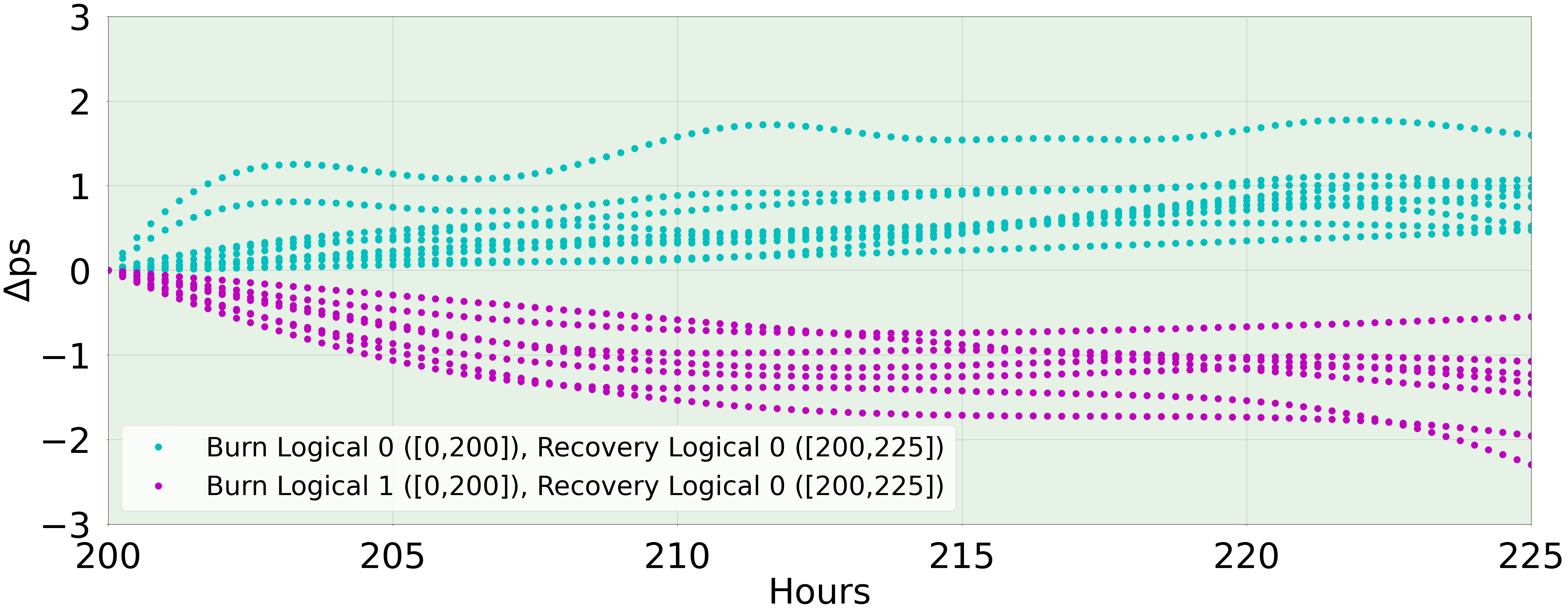}
    \vspace{-18px}
    \caption{10000 ps Routes}
    \label{fig/f1_experiment_2_1}
\end{subfigure}
\vspace{-5px}
\caption{
\textbf{Experiment 3 (Cloud Environment):} Four sets of 16 routes with different delays are initialized on an AWS F1 FPGA. A random constant burn value $X$ is conditioned into the routes for 200 hours. The attacker gains control of the FPGA and instantiates TDC sensors to measure the timing delay of routes that previously held sensitive data. The timing difference between the falling and rising transition is plotted for the following 25-hour period---revealing the BTI recovery. This enables an attacker to execute on \textbf{Threat Model 2} and recover \textbf{Type 2} data (user data).
}
\label{fig/f1_experiment_2}
\end{figure}

Figure~\ref{fig/f1_experiment_2} shows the 25-hour recovery period. Note that the graph starts at hour 200 after the burn-in. We have no data about the FPGA before that point. 
Burn 0 values are shown in cyan burn 1 values in magenta. 
The purple routes that previously held logical 1 immediately begin to decrease in relation to the cyan routes that remain at logical 0 the entire time.
The purple logical 1 routes are undergoing Burn 1 BTI recovery, which Experiment 1 showed was much more dramatic than Burn 0 recovery.

We do not observe the same magnitude and clarity of divergence of burn-in or recovery on AWS F1 as the ZCU102. 
This is likely due to more complex cloud operating factors, including non-constant temperature, FPGA age, and other computations simultaneously running on the AWS F1 system. 

Despite these differences, the result is consistent with the elasticity we observed on the ZCU102.

We show an attacker can extract Type B Data demonstrating \textbf{Threat Model 2} on a cloud FPGA.
An attacker generates a \textit{Measure} design which maps TDCs to the routes which contained \textbf{Type B} (User Data) in a targeted design.
After gaining access to an AWS F1 FPGA device running the target design that was previously loaded for tens of hours, the attacker recovers previous user data by measuring the timing behavior of the routes over time.

\section{Related Work}
\label{section/related_work}
Our attack is a \textit{single-tenant temporal side channel} --  state within the FPGA that is not wiped correctly, or it is impossible to remove between subsequent users~\cite{bobda2022future}. It is common to ``wipe'' the FPGA device between successive users~\cite{leeser2021fpgas} as a security precaution. \textbf{Our approach subverts these efforts} as it measures analog remanence that remains even after wiping. We show that our data recovery techniques work even after performing the wiping done by AWS. It is impossible to mitigate burn-in risk via a logical erasure of the device because burn-in is a fundamental characteristic of the device transistors that reflects previous logical values. 

Tian et al.~\cite{tian2019temporal} demonstrate a single-tenant temporal covert channel. They use ring oscillators to heat the FPGA (transmitter) and detect temperature (receiver).  They can transmit hundreds of bits over a few minutes on cloud FPGA at Texas Advanced Computing Center using Microsoft Catapult hardware. To make their covert channel, the FPGA transmitter and receiver must alternatively obtain and release the same FPGA, which is possible but very difficult in other cloud infrastructures (e.g., AWS).  
Using temperature as a side channel requires the user to get on the FPGA quickly; temperature effects are short-term, e.g., the cloud FPGAs return to ambient temperatures within a few minutes~\cite{tian2019temporal}.
Finally, BTI effects are a more pernicious temporal channel. Instead of measuring the tertiary effects of computation or a covert channel, it is a direct measurement of a previous user or proprietary design data. It can last hundreds of hours, as we have shown in our results.

Zick et al.~\cite{zick2014high} demonstrate a single-tenant temporal side channel on a local FPGA by recovering previous user data stored in LUT SRAMs. Their experiment has a burn-in period of 922 hours at high temperatures to induce burn-in. Then, the FPGA sat powered off for several weeks. Their experiments were performed on a local Xilinx Kintex-7 KC705 development board. Unfortunately, their experimental requirements are incompatible with the cloud FPGA attack model. They use a highly precise, off-chip oscillator to enhance the on-chip TDC sensor timing resolution. This results in femtosecond-level timing precision. Such precision is impossible on cloud FPGA TDC sensors since an attacker cannot use off-chip components. On-chip TDCs operate at approximately 10 ps precision on the UltraScale+, so it is an order of magnitude difference with their sensor. 
They perform recovery of data stored in FPGA LUTs (SRAM) and specifically target transistors in the output buffers of the SRAM bits. We ruled out the examination of this resource since their burn-in effects are too subtle to measure with cloud FPGA sensors, which is why they required femtosecond precision. We target FPGA programmable routing. We show that our attacks are deployable on cloud FPGAs (AWS F1 instances). 

A significant body of prior work uses ring oscillators (RO)-based sensors to measure long-term FPGA BTI effects~\cite{6645584, amouri2014aging, stott2010degradation}. RO sensors build a combinatorial loop through a tested component and an inverter.
The oscillation frequency through the loop reflects the time taken for the signal to propagate through that tested component, which changes due to BTI effects. While ROs measure BTI effects, they have two significant limitations.
First, ROs have a single variable output---the frequency of oscillation – that integrates the propagation speed through the NMOS and PMOS transistors. This is an essential factor as BTI stresses PMOS vs. NMOS transistors differently. Our TDC sensor can separate the differences in BTI stress on PMOS and NMOS. We use this ability to differentiate between BTI degradation. 
Second, ROs are often not allowed on cloud FPGAs. ROs use combinatorial loops, which violate the design rule checks and can be detected~\cite{krautter2019mitigating, la2020fpgadefender}. Cloud FPGA providers can disallow designs that contain self-oscillating circuits, e.g., as is done by AWS. Our TDC-based sensor is more challenging to detect since it uses computational structures that are common and many FPGA designs. It was implemented on an AWS F1 instance. Thus, it passes AWS design rule checks. 

Previous works have recovered SRAM user data on recycled ICs~\cite{https://doi.org/10.48550/arxiv.2208.02883, trimberger2013system, cakir20126t}.
Even though SRAMs are a form of volatile memory, where logical data is lost on power-off, an imprint is left behind and is recoverable. These techniques rely on measuring the statistical power-on state of SRAM bits. They assume a different threat model, e.g., requiring physical access to the chip.

\section{Mitigations}
\label{section/mitigations}
This paper demonstrated that \textbf{Threat Model 1} and \textbf{Threat Model 2} are exploitable in cloud systems. A determined attacker could build more precise sensors to measure BTI on shorter routes with shorter burn-in periods. As a result, users should take precautions to manage sensitive data to mitigate burn-in effects, cloud FPGA providers should look to enforce stronger temporal boundaries between users, and FPGA manufacturers should consider architectural solutions to mitigate BTI.

\subsection{User Mitigations}
The cloud FPGA user should not allow sensitive data to sit unchanged on the FPGA for long periods to avoid burn-in remnants that the following user could discover. When sensitive data must statically persist for long periods, the user should consider methods to mitigate its burn-in effects. 

Techniques that periodically change sensitive data would reduce burn-in. For example, the data could be inverted at predetermined periods (e.g., every hour). Or it could be deterministically shuffled at the source and unshuffled at the receiver. Such data transformation approaches reduce the burn-in effects across the route at the expense of increasing the design complexity. Other ideas related to FPGA wear leveling~\cite{stott2011improving} would likely reduce the burn-in effects but need to be verified.

If there are natural breaks in computation, the user could move between different FPGAs in the cloud. A new device should be checked out from the cloud provider, the application moved between FPGAs, and the burn-in would start fresh on the new FPGA. This would increase user design complexity, e.g., by requiring a robust process to stop, move application data, and restart an FPGA instance. This adds risks related to data corruption when moving between FPGAs.

The user should strive to make routes that hold sensitive data as short as possible. The longer the route, the more transistors affected by burn-in, and the larger the burn-in effects. Consequently, as we have shown in our results, shorter routes are a more secure FPGA design pattern. 

FPGA physical design tools generally attempt to make routes as short as possible. Hand-placed routes for sensitive information could produce better results. Physical design tools often focus on routes on the critical path, often at the expense of other routes. The ability to specify that the physical design tools minimize sensitive routes would reduce vulnerability to pentimento-style attacks.  

The user or design tools could place sensitive routes in a manner that makes them difficult to connect to a BTI sensor. The input to the route under test must be connected to the transition generator. The output of the route under test is connected to a TDC sensor. Placing the inputs or outputs in locations that make these connections challenging would make it more challenging to extract BTI information from that target route. 

Verification tools could analyze the design or bitstream for sensitive data residing on long routes. The ability to provide reports about the route lengths of the sensitive information would allow hardware security verification engineers to better assess their data vulnerabilities  w.r.t. to a pentimento attack. Providing a more precised measure of protection (e.g., vulnerability metric) enables  even stronger hardware security verification. 

Key rotation is common in cryptography~\cite{boneh2013key, everspaugh2017key} and could be employed on cloud FPGAs. This is not always possible, especially if data needs to be embedded into the RTL directly, e.g., in random netlist constants as found in the OpenTitan. 

Key masking~\cite{akkar2001implementation, golic2002multiplicative, mangard2005side} could help reduce the number and lengths of routes that hold a cryptographic key. Masking is specific to cryptographic algorithms and may not be feasible for other types of sensitive data.

The design could use partial reconfiguration to move the sensitive information -- its storage and computation units -- to different locations of the chip. This would act as a form of wear leveling. It lessens the burn-in effect at any one physical location. Yet, it also spreads the burn-in over more areas, which could potentially make it easier to exploit the information.  

A cloud FPGA user could mitigate the BTI remnants by erasing their design and holding on to the instance for some time before relinquishing it back into the user pool. The tenant could invert the values of the sensitive routes to speed up the recovery and thus limit the remaining BTI signal. Or they could perform some other actions (perhaps toggling the routes). This costs the user money commensurate with the time they deemed sufficient to erase BTI effects. 

\subsection{Cloud Provider Mitigations}
The primary issue cloud providers could hope to resolve is the rapid reallocation of FPGA devices once a user relinquishes them. The cloud provider could implement launch rate controls, by withholding devices after they are returned, for days, weeks, or longer to mitigate the ability to recover the burn-in. This would push the mitigation onto the cloud provider rather than the cloud user.

The cloud provider can attempt to combat the accelerators of the BTI effect: higher voltage and temperature. Some FPGAs that operate at different voltages and use a lower voltage would reduce the burn-in effects. Similarly, higher temperatures exacerbate burn-in. Temperature can be managed to some extent, but it would be very challenging to control the on-chip temperature to the point where an attacker can no longer observe BTI. Furthermore, cloud providers are already incentivized to control voltage and temperature to reduce FPGA power consumption and aging. 

\subsection{FPGA Manufacturer Mitigations}
FPGA manufacturers can attempt to mitigate FPGA BTI effects. BTI mitigations are already commonly considered to increase reliability.  It is unlikely that FPGA manufacturers will be able to eliminate BTI, especially at advanced design nodes. BTI effects are more negligible at less advanced process nodes; thus, falling back on older technology would be a potential mitigation. The performance and power benefits of advanced nodes are likely too much to sacrifice for cloud providers and users.

Manufacturers can help reduce BTI through voltage and temperature mitigations; however, this is already a primary directive due to their negative influence on power consumption. Thus, it is unlikely these mitigations will advance at a faster pace.  FPGA manufacturers could consider more advanced dynamic voltage scaling techniques to allow users to mitigate BTI selectively. This adds complexity to the design, which increases costs.
\section{Conclusion}
\label{section/conclusion}
We find that a remote attacker can recover “FPGA
pentimentos” – long-removed secret data belonging to a prior
user or proprietary design image on a cloud FPGA. Just as a
pentimento of a painting can be exposed via infrared imaging,
FPGA pentimentos can be exposed via signal timing sensors
instantiated on a remote cloud FPGA.
The sensitive data constituting an FPGA pentimento is
imprinted to the device through bias temperature instability
effects on the underlying transistors. We demonstrate how this
slight degradation can be measured using a time-to-digital
converter when an adversary programs one into the target
cloud FPGA. This technique allows an attacker to ascertain
previously safe information, after it is no longer explicitly
present, on cloud FPGAs. Notably, it can allow an attacker
to (1) extract proprietary details or keys from an encrypted
FPGA design image available on the AWS marketplace and (2)
recover information from a previous user of a cloud-FPGA.
Both threat models are experimentally validated on the AWS
F1 platform.

\bibliographystyle{plain}
\bibliography{bibliography}

\end{document}